\documentclass[paper=letter, fontsize=10pt,leqno]{scrartcl}
\usepackage[body={5.35in,7.5in}]{geometry} 

\usepackage[english]{babel}				
\usepackage{amsmath,tabu}				
\usepackage{amsfonts}
\usepackage{amsthm}
\usepackage{latexsym}
\usepackage{hyperref}

 \usepackage{verbatim}
\usepackage[pdftex]{graphicx}				
\usepackage{url}
\usepackage{amssymb}
\usepackage{bbm}
\usepackage{MnSymbol}

\usepackage[pdftex]{color}
\usepackage{eucal}

\usepackage{amscd}
\usepackage{calrsfs}
\usepackage[format=plain]{caption}
\usepackage{wrapfig}
 \usepackage{float}
 
 \usepackage{longtable}
 \usepackage{booktabs}
 \usepackage{array}
 
 \usepackage{bigints}

\usepackage{sectsty}					
\allsectionsfont{\centering \normalfont\scshape}	

\usepackage{mathtools}
\DeclareCaptionStyle{ruled}{labelfont=bf,labelsep=space}
\usepackage[]{forest}

\newtheorem{theorem}{Theorem}

\newtheorem{proposition}[theorem]{Proposition}
\newtheorem{lemma}[theorem]{Lemma}
\newtheorem{corollary}[theorem]{Corollary}

\newtheorem*{proposition*}{Proposition}

\title{
		\vspace{-1in} 	
		\usefont{OT1}{bch}{b}{n}
		\normalfont \normalsize \textsc{} \\ [25pt]
		\huge  Spin-geodesic systems
}

\hypersetup{hidelinks}

 \date{\today}
 \author{\normalfont \large 
  R. Feres\footnote{\scriptsize Department of Mathematics and Statistics, Washington University, Campus Box 1146, St. Louis, MO 63130},
\ Z. Hu\footnotemark[1]
\ 
}

\begin{document}

\maketitle

\begin{abstract}
\begin{center}
 Abstract \end{center}
{\small  
We introduce and study a class of geometric dynamical systems on Riemannian manifolds called spin-geodesic systems, in which the motion of a point is coupled to an internal classical spin through a symmetric endomorphism field $B$, the spin-coupling field. These systems may be viewed as $B$-dependent deformations of geodesic motion and have connections with frame flows, magnetic geodesic flows, and nonholonomic rolling systems. We show that every spin-geodesic system admits an equivalent interpretation as a no-slip rolling system in an abstract ambient geometry; when the associated metric connection, determined by $B$, is flat, this construction reduces, under the usual integrability hypotheses, to ordinary rolling on a Euclidean hypersurface. Our second main result is a Liouville theorem: for an arbitrary symmetric spin-coupling field B, the spin-geodesic flow preserves the canonical Liouville volume on its mechanical phase space. This holds even though the flow is  not  symplectic unless $B$ is identically zero.
}
\end{abstract}

\section{Introduction}
Spin-geodesic systems constitute a class of dynamical systems of geometric-mechanical nature whose configuration manifold 
is the orthonormal frame bundle $\mathcal{F}=\text{SO}(M)$ of an oriented Riemannian manifold
  $(M, \langle\cdot, \cdot\rangle)$. Thus a trajectory of such a system has, in addition to a translational part represented by a point in $M$, 
  internal degrees of freedom---the fibers of the principal bundle---defining an orientation in space whose rate of change will be referred to as (classical mechanical) {\em spin}.
  Translational and rotational  motions interact via  the choice of a {\em spin-coupling field} $B$---a  symmetric  section of the vector bundle $\text{End}(TM)$. Different choices of the spin-coupling field  lead to a variety of dynamical systems of differential geometric/mechanical type such as geodesic flows, magnetic geodesic flows, rolling flows (the latter are non-holonomic systems to be defined shortly), and systems that share an affinity with frame flows.
  
  The system's phase space   is the tangent bundle $T\mathcal{F}$, and the \emph{reduced phase space} is $\mathcal{A}:=T\mathcal{F}/\text{SO}(d)$, where $d$ is the dimension of $M$ and $\text{SO}(d)$ is the special orthogonal group in dimension $d$. This quotient is well-known to be an Atiyah (Lie) algebroid. (Basic facts concerning Atiyah algebroids and groupoids will be reviewed later.) $\mathcal{A}$ is a vector bundle over $M$ which, in the presence of the Levi-Civita connection  for the given   Riemannian metric on $M$, splits as a direct sum
  \[\mathcal{A}\cong TM\oplus \mathfrak{so}(TM),\]
  where $\mathfrak{so}(TM)$ is the bundle of Lie algebras over $M$ whose fiber over $x\in M$ is the space of skew-symmetric endomorphisms of $T_xM$. The corresponding Atiyah groupoid  is the quotient $\mathcal{G}=(\mathcal{F}\times \mathcal{F})/\text{SO}(d)$ of the product $\mathcal{F}\times \mathcal{F}$ by the diagonal action of the orthogonal group. Its arrows are linear isometries $T_xM\rightarrow T_yM$; the pair $(x,y)\in M\times M$  (the groupoid associated to the algebroid $TM$) defines a finite translation in $M$ 
  and the linear isometry is a relative rotation. Thus a motion of the system consists of  translation along a curve $x(t)$ on $M$ and
  a  family of rotations $R(t): T_{x(0)}M\rightarrow T_{x(t)}M$ parametrized by time.

 The equations of motion of the spin-geodesic system are determined from the data $(M, \langle\cdot,\cdot\rangle, B)$. We indicate the system
 by this triple.  
Before defining  the dynamical equations, let us fix some notation. Let $\nabla=\nabla^M$ denote  the Levi-Civita connection on   $TM$ as well as   the induced connection on endomorphism tensor fields, hence, in particular, on sections of $\mathfrak{so}(TM)$.
The covariant derivative of a vector or tensor field along a parametrized differentiable curve will be indicated by $\frac{\nabla}{dt}$, rather than by the more common $\frac{D}{dt}$. (This is needed so as to distinguish the different connections that will appear throughout our discussion.) 
Let  $S$ denote elements of $\text{SO}(T_xM)$,  $x$ elements of $M$; we typically write sections of $TM$ and  $\mathfrak{so}(TM)$ as   $v$ and $W$, respectively. 
The spin-geodesic system can now be written as
\begin{equation}\label{eq:spin_geodesic_system}
\dot{x}=v, \ \ \ \frac{\nabla v}{dt}= -WBv, \ \ \ \frac{\nabla W}{dt} = (Bv)\wedge v,
\end{equation}
to which we add the reconstruction equation for the rotation part:
\begin{equation}\label{eq:reconstruction_equations}
\frac{\nabla S}{dt} = SW.
\end{equation}
The third equation in  (\ref{eq:spin_geodesic_system}) uses   the (generalized) cross-product, $\wedge: T_xM\times T_xM\rightarrow \mathfrak{so}(T_xM),$ defined as
\begin{equation}\label{eq:generalized_cross_product}
(u\wedge v) w :=\langle u, w\rangle_x v-\langle v,w\rangle_x u.
\end{equation}
When we later consider the example of rolling systems, it  should be borne in mind that the quantities involved in \eqref{eq:spin_geodesic_system} have the following physical dimensions: components of $W$ are velocities (angular velocities multiplied by a particle radius $r$ and a dimensionless parameter $\eta$ derived from the scalar moment of inertia associated to a rotationally symmetric mass distribution); $v$ is naturally a velocity, and components of  $B$ are the reciprocal of distance (due to its later  interpretation as $\eta$ times a shape operator.)

We define a Riemannian metric on $\mathcal{A}$ as follows: given $(v_1, W_1), (v_2, W_2)\in \mathcal{A}_x$ (the fiber  of $\mathcal{A}$ above $x\in M$), then
\begin{equation}\label{eq:metric_on_A}
\left\langle (v_1, W_1), (v_2, W_2) \right\rangle_x = \langle v_1, v_2\rangle_x +\frac12 \text{Tr}\left(W_1W_2^\intercal\right).
\end{equation}
  With respect to this (kinetic energy) Riemannian metric, the spin-geodesic system is energy-preserving for any choice of the symmetric field $B$. One of the main results of the paper is that, for all choices of $B$, the spin-geodesic flow leaves invariant a canonical (Liouville) volume form, even though it does not preserve the canonical symplectic form unless $B= 0$ identically. This    reflects an underlying non-holonomic aspect of the spin-geodesic system that will become apparent in  its more explicitly mechanical manifestation (rolling systems).
  
Spin-geodesic systems are related to various well-established dynamical systems of  differential geometric kind. When $B=0$,  trajectories $x(t)$ are geodesic paths in  $M$ and the spin $W$ is parallel transported along this base path. In this sense, this special case resembles frame flows, although it is the frame spinning, rather than the frame  itself, which is parallel transported. For this reason,   we may regard  the flow of the new systems   as $B$-deformations of geodesic flows. It is this perspective  we wish to emphasize by the term ``spin-geodesic.''

 \begin{figure}[htbp]
\begin{center}
\includegraphics[width=4.0in]{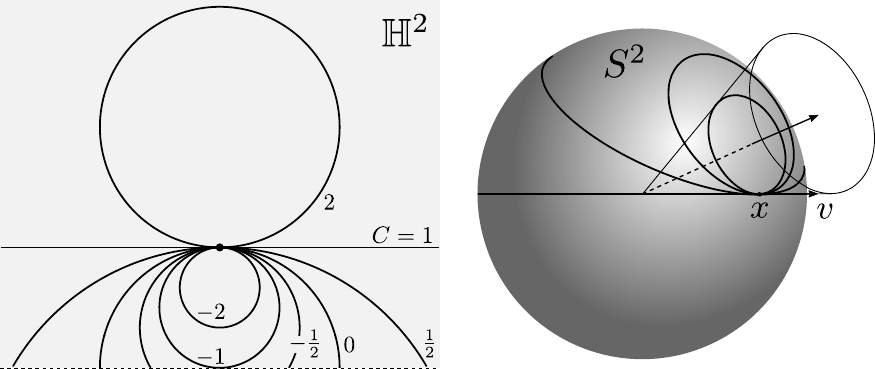}\ \ 
\caption{{\small  When $M$ has dimension $2$ and $B$ is a  scalar operator (a smooth function times the identity), the spin-geodesic system is a magnetic geodesic system. The   examples in this figure assume $B=CI$, $C\in \mathbb{R}$. Left: $M$ is the hyperbolic plane and the indicated numbers are values of $C$.
These  are: geodesics for  $C=0$, horocycles for $|C|=1$, hypercycles for $|C|<1$, and hyperbolic circles for $|C|>1$.  All have initial conditions $x_0=(0,1)$ and $v_0=(1,0)$. Right:  $M$ is a sphere. Solutions to the spin-geodesic equations are the magnetic geodesic trajectories on the sphere under a magnetic monopole field \cite{CFZ21}. Their projections  in $S^2$ are the intersection of the sphere with cones.}}\label{fig:curves_constant_geo_curvature}
\end{center}
\end{figure}  

When $B_x=f(x) I$,
  the dynamical equations reduce to 
  \[\frac{\nabla v}{dt}=-f Wv, \ \ \ \frac{\nabla W}{dt}=0.\]
  In particular, when $d=2$, the bundle of Lie algebras is a trivial rank-$1$ vector bundle and we may write $W=sJ$, where $J$ is the complex structure (that is, $J_x:T_xM\rightarrow T_xM$ is positive rotation by $\pi/2$ for each $x\in M$). The previous  pair of equations then becomes
  \[\frac{\nabla v}{dt} = -f s Jv, \ \ \ \dot{s}=0,\] 
  thus the spin $s$ is constant.  In this case, the spin-geodesic system becomes a  magnetic geodesic system on a surface  with  possibly nonuniform magnetic intensity $sf$. If $f$ is constant, we have the equation
$\frac{\nabla \dot{x}}{dt}=C J\dot{x}$, where $C$ is a real constant.
  When $M$ is, for example, the hyperbolic plane,  solutions are known 
  to be curves of constant geodesic curvature.  On a sphere, solutions are the trajectories of a charged particle under the   field of a magnetic monopole. See Figure \ref{fig:curves_constant_geo_curvature}.

When $M$ is a hypersurface in Euclidean space and $B$ is, up to a positive multiplicative constant, the shape operator of the isometric immersion,
the spin-geodesic system gives the rolling systems studied in \cite{CCCF20,CFZ21,CFH26}. We will have much more to say about this manifestation of equations \eqref{eq:spin_geodesic_system} but, briefly, rolling systems consist of a ball with rotationally symmetric mass distribution rolling without slip over a submanifold of $\mathbb{R}^{d+1}$ (not necessarily a hypersurface and possibly with boundary). The $d$-dimensional manifold $M$ is, in this case, the locus of centers of the rolling ball and $B$ is the shape operator of the immersion-induced metric on $M$ multiplied by
a quantity derived from the ball's moment of inertia. 
The example on the right-hand side of Figure \ref{fig:curves_constant_geo_curvature}, in addition to its interpretation as a magnetic-geodesic system, 
  can  be regarded as modeling the motion of a ball rolling without slip over a fixed point at the origin of $\mathbb{R}^3$.

The paper contains two  main results.  The first is Theorem \ref{theor:abstract_rolling_equations}. It shows that spin-geodesic systems are equivalent
to what we call \emph{abstract rolling systems}, or rolling systems on \emph{abstract ambient geometries}. The main idea is as follows. On the Riemannian vector bundle  $E=TM\oplus\, \mathbb{R}\nu$  over $M$ of rank $d+1$,
where $\nu$ is a nonvanishing section of a trivial real line bundle, 
we introduce a Riemannian metric $\langle\cdot, \cdot\rangle^E$ making $\nu$ a unit vector orthogonal to $TM$, and
  a metric connection $D^\mathbb{S}$ where $\mathbb{S}=\eta^{-1}B$ for a choice of positive constant $\eta<1$. This constant is  interpreted as a moment of inertia parameter.  We call $\left(E,\langle\cdot,\cdot\rangle^E,D^\mathbb{S}\right)$  the \emph{abstract ambient geometry}. The spin-geodesic equations can then be expressed as rolling equations with a no-slip nonholonomic constraint. This is in the spirit of Cartan geometries. (See appendix B of \cite{Sharpe}; in Cartan geometries one adds a \emph{no-twist} condition to uniquely derive the development of a model geometry along a path over the manifold. For the dynamical rolling, it is natural to drop the no-twist constraint. It would impose an
  additional constraining torque,  which is absent from the mechanical model considered here.) When the connection $D^{\mathbb{S}}$ is flat, 
 $M$ admits an isometric immersion as a Euclidean hypersurface with shape operator $\mathbb{S}$ and
the spin-geodesic equations become concrete rolling equations.

The second main result (Theorems   \ref{theor:main_vol_invariance} and  \ref{theor:volume_invariant_levels}) is that  spin-geodesic flows leave invariant the canonical (Liouville) volume form on
an associated phase-space manifold, for all spin-coupling fields $B$. The same holds for the restriction of the flow to constant energy hypersurfaces. 
This volume invariance should be contrasted with Theorem \ref{theor:not_symplectic}, which shows that the symplectic form itself is not preserved unless $B$ is identically $0$.  That such mechanical/geometric systems are not  Hamiltonian with respect to the canonical symplectic structure considered here is to be expected due to their underlying nonholonomic nature; the preservation of the canonical volume form for all choices of $B$, however, should be of note.

Having a Liouville theorem for spin-geodesic systems opens up a path for developing  their ergodic theory along the lines  as   for geodesic flows.
Concerning very specific examples and dynamical properties of spin-geodesic systems in their  rolling system manifestation, and the connection with what has been called \emph{no-slip billiard} systems, we again point   to  \cite{CCCF20,CFZ21,CFH26}.

The plan of the paper is as follows. In section \ref{sec:Equivalence} we prove the first main result, concerning the equivalence of spin-geodesic systems and rolling systems in abstract ambient geometries. We begin the section with a review of concrete rolling systems on submanifolds of $\mathbb{R}^{d+1}$, including  both the free constrained motion and motion under external forces. The equivalence theorem, including  external forces, is given in subsection \ref{sec:Rolling_abstract}.   The Liouville theorem for spin-geodesic flows is developed in section \ref{sec:Liouville_theorem}. We begin that section by clarifying a number of closely related spaces on which the spin-geodesic dynamics may be 
studied and how the canonical invariant volume is described in each space. One space, in which the spin-geodesic equations take on their simplest form is the Atiyah algebroid $\mathcal{A}$. The other is the space of rotations $N$. The latter is particularly convenient since the Liouville volume form becomes on it  the volume arising from the symplectic form on $T^*N$. After comparing the two settings, we proceed  to work mostly on the Riemannian manifold $N$ with its induced Kaluza-Klein metric.  It turns out that $TN$ admits two natural metric connections: its Levi-Civita connection $\nabla^N$ and the  connection $D$ induced from the Levi-Civita connection on $M$. The latter has a non-zero  torsion tensor which reflects the curvature of $M$. The spin-geodesic equations expressed in terms of $\nabla^N$ incorporate an additional force term of geometric origin that has now to be taken into account. After these Riemannian geometry preliminaries, we finally prove the Liouville theorem for spin-geodesic systems in section \ref{sec:volume_invariance_final}. In the final section \ref{secion:non-symplectic} we show
that the system is not symplectic unless $B$ is identically zero.

  \section{Equivalence of  spin-geodesic and rolling systems}\label{sec:Equivalence}
  The main interpretation of a spin-geodesic system is as a   rolling system in an abstract ambient geometry built from the data $(M,\langle\cdot,\cdot\rangle, B)$. This interpretation is based on
  \cite{CFZ21} and \cite{CFH26}. We briefly review concrete rolling systems, then proceed to discuss their abstract general form and their close connection with spin-geodesic systems.

  \subsection{Rolling on submanifolds of $\mathbb{R}^{d+1}$}
\begin{figure}[htbp]
\begin{center}
\includegraphics[width=3.0in]{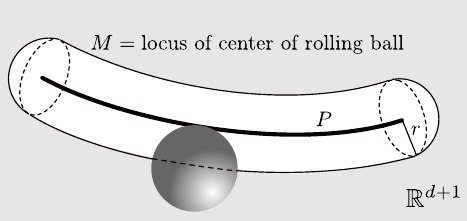}\ \ 
\caption{{\small A ball of radius $r$ rolls over a submanifold $P$ of $\mathbb{R}^{d+1}$ satisfying the no-slip constraint. The hypersurface consisting of the locus of centers of the ball is denoted $M$.  }}\label{fig:rolling_figure}
\end{center}
\end{figure}

For a given   submanifold $P\subseteq \mathbb{R}^{d+1}$ of positive codimension and possibly nonempty boundary, let $M$ be the boundary of the tubular neighborhood of $P$  of radius $r$. See Figure \ref{fig:rolling_figure}. For small enough $r$, $M$ is a differentiable hypersurface without boundary. We regard $M$ as the set of centers of a ball of radius $r$ which remains in tangential contact with $P$.
The configuration manifold of the rolling-ball system is the submanifold of the Euclidean group \[\mathcal{M}:=\left\{(x, A)\in \text{SE}(d+1)=\mathbb{R}^{d+1}\rtimes \text{SO}(d+1): x\in M\right\}.\] 
  A motion of the system is a differentiable path $q(t)=(x(t), A(t))\in \mathcal{M}$; we express the velocity $\dot{q}(t)$ as
  \[\dot{q}(t)=\left(\dot{x}(t), U(t)A(t)\right), \ \ \ U(t) = \dot{A}(t)A(t)^{-1}\in \mathfrak{so}(d+1),\] where $U(t)$ is the angular velocity of the rolling ball and $\dot{x}(t)\in T_{x(t)}M$ is
  the velocity of the center.

Denote by $\nu(x)$ the outward pointing unit vector field to $M$.  For a state $\xi\in T_q\mathcal{M}$, $q=(x,A)$,  the velocity of the point of contact of the ball with $P$ is
\[V(q,\xi) =u_\xi - rU_\xi\nu(x), \ \ \ \xi=(u_\xi, U_\xi A).\]
  The no-slip condition is the constraint  that $V(q,\xi)=0$. This condition defines a vector subbundle $\mathfrak{R}$ of $T\mathcal{M}$, called
  the \emph{rolling bundle} in \cite{CFZ21}. Thus
  \[\mathfrak{R}_q:=\left\{(u,UA)\in T_q\mathcal{M}: u = rU\nu(x)\right\}.\]
 When  $\dot{q}(t)\in \mathfrak{R}_{q(t)}$ for all $t$,  we say that the motion $q(t)\in \mathcal{M}$ satisfies the \emph{no-slip} (nonholonomic) constraint.

  Given a rotationally symmetric mass distribution on the rolling ball of  total mass $m$, its matrix of centered  second moments
  is scalar. We write it as $\lambda I$, where 
  \[\lambda = \frac12 m \left(r\gamma\right)^2,\]
  in which $\gamma$ is a dimensionless quantity. The system's kinetic energy is shown to be the field of quadratic forms on the tangent bundle of $\mathcal{M}$ whose polarization is the inner product
\begin{equation} \label{eq:kinetic_Riemannian_metric}\left\langle \xi, \zeta\right\rangle_q :=m\left\{\langle u_\xi,  u_\zeta\rangle_x  + \frac{(r\gamma)^2}{2} \text{Tr}\left(U_\xi U_\zeta^\intercal\right)\right\}.\end{equation}
   Here $\langle u_\xi,  u_\zeta\rangle$ is the standard (dot) inner-product in $\mathbb{R}^{d+1}$ restricted to tangent spaces of $M$.
   The field of inner products in  \eqref{eq:kinetic_Riemannian_metric} defines a Riemannian metric on $\mathcal{M}$.  
   Relative to this metric, the orthogonal complement of the rolling bundle is  
   \[\mathfrak{R}_q^\perp =\left\{\left(w, \frac1{r\gamma^2}w\wedge \nu(x)\right): w\in T_xM\right\}.\]
Newton's equation for the rolling motion is obtained by the Lagrange-d'Alembert principle:  if $\nabla$ is the Levi-Civita connection
for the above Riemannian metric on $\mathcal{M}$, then $q(t)$ is a motion satisfying the no-slip constraint if
\[m\frac{\nabla \dot{q}}{dt} =N(q,\dot{q})\]
where $N(q(t),\dot{q}(t))\in \mathfrak{R}_{q(t)}^\perp$ and $\dot{q}(t)\in \mathfrak{R}_{q(t)}$ for all $t$.
\begin{proposition}[\cite{CFZ21}]\label{prop:rolling_eq_U}
Let $q(t)=(x(t), A(t))$ be the motion of the rolling ball with rotationally symmetric mass distribution and moment of inertia parameter $\gamma$, subject to the nonholonomic constraint defined by the rolling distribution $\mathfrak{R}\subseteq T\mathcal{M}$. Then 
\[\dot{U}=-\frac{r}{1+\gamma^2}\left(U \mathbb{S}_xU\nu(x)\right)\wedge \nu(x),\]
subject to the constraint equation $\dot{x}=u=rU\nu(x).$
\end{proposition}

In order to see that the equations of Proposition \ref{prop:rolling_eq_U} can be rewritten as a spin-geodesic system of equations, let us first observe that any $U\in \mathfrak{so}(d+1)$ can be decomposed as 
\begin{equation}\label{eq:ortho_dec_skew}U=\Pi_x U\Pi_x +\nu(x)\wedge U\nu(x),\end{equation}
where $\Pi_x$ is the orthogonal projection from $\mathbb{R}^{d+1}$ to $T_xM$. Under the rolling constraint, and further defining
\begin{equation}\label{eq:definition_W}W:=r\eta \Pi_x U \Pi_x,\ \ \ \eta:=\frac{\gamma}{\sqrt{1+\gamma^2}}\end{equation} we have
\[U = \frac1r \left(\eta^{-1} W +\nu(x) \wedge u\right).\]
In this form, we can express the equations of motion in terms of $W$ and $u$ and eliminate the nonholonomic constraint equation. The resulting equations are as follows. We call $W$ the rolling ball's  {\em tangential spin}.
\begin{theorem}[Rolling equations, \cite{CFZ21}]\label{theor:rolling_equations_W}
The rolling motion with hypersurface (of ball centers) $M$ under the no-slip constraint satisfies  
\[\frac{\nabla u}{dt}= -\eta W \mathbb{S}_x u, \ \ \ \frac{\nabla W}{dt}=\eta\left(\mathbb{S}_x u\right)\wedge u, \ \ \ \dot{x}=u,\]
where $\nabla$ is the Levi-Civita connection of the hypersurface $M$ and $\mathbb{S}$ is its shape operator.
This is a spin-geodesic system $(M,\langle\cdot,\cdot\rangle, B)$ where the spin-coupling field is $B=\eta\mathbb{S}$.
\end{theorem}

\subsection{Remark about external forces}
Let $\mathcal{B}$ denote the ball of radius $r$ centered at the origin of $\mathbb{R}^{d+1}$. Denote by $\mu$ the rotationally symmetric mass distribution on
$\mathcal{B}$ having total mass $m$. 
 Suppose that an external force  distribution is defined by the vector field $\varphi(x)$  on a domain containing the possible configurations of the ball,
 so that at each point $b\in \mathcal{B}$ and each configuration $q=(x,A)$, an element of mass $d\mu(b)$ is subject to a force
 $\varphi(Ab+x)\, d\mu(b)$. Given  $\xi\in T_q\mathcal{M}$, let $V_\xi(b):=U_\xi Ab+u_\xi$ denote the velocity of the material point $b$ in the state $(q,\xi)$.  From $\varphi$ we obtain the {\em power functional} $\hat{\varphi}_q$, a $1$-form on $\mathcal{M}$ given by
 \[\hat{\varphi}_q(\xi):=\int_\mathcal{B} V_\xi(b)\cdot \varphi(Ab+x)\, d\mu(b).\]
  The dual of this $1$-form relative to the Riemannian metric on $\mathcal{M}$ is the vector field (depending only on $x$ and not in  $A$)
  which we call a {\em wrench}: \[(\psi(x), \Psi(x))\in T_xM\oplus  \mathfrak{so}(d+1),\] so that
  \[\hat{\varphi}_q(\xi) =\left\langle (u_\xi, U_\xi),(\psi(x), \Psi(x)) \right\rangle_q. \]
    The wrench components are
    \[\psi(x)=\int_B \varphi(b+x)\frac{d\mu(b)}{m}, \ \ \  \Psi(x)=\frac1{(r\gamma)^2} \int_B b\wedge \varphi(b+x) \frac{d\mu(b)}{m}.\]
    The above description of $\varphi(x)$ as a force density  relative to the mass distribution measure $\mu$
    is easily generalized by assuming a singular field of forces; this may describe, for example, forces acting on a finite set of material points of $\mathcal{B}$.

   The same Lagrange-d'Alembert  argument used to obtain the equations for the constrained motion  in the absence of forces
   can be used to prove the following.  
   \begin{theorem}[Rolling equations under external forces, \cite{CFH26}]
   With the notations just given, let us write 
   \[f:=r\eta^2 \Psi \nu +(1-\eta^2)\Pi\psi, \ \ \ \tau:=r\eta\Pi\Psi \Pi.\]
   Then the rolling equations become
   \begin{equation}\label{eq:spin_geodesic_force}\frac{\nabla u}{dt}= -W B_x u+f, \ \ \ \frac{\nabla W}{dt}= \left(B_x u\right)\wedge u +\tau, \ \ \ \dot{x}=u,\end{equation}
   where $B=\eta\, \mathbb{S}$.
   \end{theorem}

  \subsection{Rolling systems in an abstract ambient geometry}\label{sec:Rolling_abstract}
  Given  a general spin-geodesic system $(M,\langle\cdot,\cdot\rangle, B)$, form the rank-$(d+1)$ vector bundle $E=TM\oplus \mathbb{R}\nu$ over $M$, where
  $\nu$ is a distinguished unit section of the trivial line bundle and the direct sum is orthogonal. We denote by $\Pi:E\rightarrow TM$ the orthogonal projection. Let us  introduce a positive constant $\eta<1$ and a field of endomorphisms $\mathbb{S}:=\eta^{-1}B$, with which we define a connection $D^\mathbb{S}$ on $E$ by the 
  following properties:
  \begin{equation}\label{eq:definition_DS}
 D^\mathbb{S}_X \nu =-\mathbb{S}X, \ \ \ 
 D^\mathbb{S}_XY = \nabla_XY +\langle X, \mathbb{S}Y\rangle \nu,
 \end{equation}
where $X,Y\in \mathfrak{X}(M)$.  We refer to  $\left(E, \nu, D^\mathbb{S}\right)$ as the  {\em abstract ambient geometry} in which $M$ formally lies   as a hypersurface with shape operator $\mathbb{S}$. (We do not yet impose the Gauss and Codazzi equations on $\mathbb{S}$, which it would satisfy if $M$ were an immersed  hypersurface in Euclidean space.) Let $\mathcal{F}^E\rightarrow M$ be the
oriented orthonormal frame bundle of $E$. Elements $\sigma\in \mathcal{F}^E$ are linear isometries, $\sigma:\mathbb{R}^{d+1}\overset{\cong} {\longrightarrow}E_x$.
Given a path $\sigma(t)\in \mathcal{F}^E$ over $x(t)\in M$  define  its {\em angular velocity}
\[U(t)=\left(\frac{D^\mathbb{S}\sigma}{dt}(t)\right)\sigma(t)^{-1}\in \mathfrak{so}(E_{x(t)})\]
and impose on $v(t)=\dot{x}(t)$ and $U(t)$ the \emph{no-slip constraint}
\begin{equation}\label{eq:abstract_constraint}v(t)=rU(t)\nu.\end{equation}
This constraint equation defines a vector subbundle $\mathfrak{R}\subseteq TM \oplus\, \mathfrak{so}(E)$, which we may continue to call the \emph{rolling subbundle}.
The positive parameter $r$ is interpreted as a particle radius.  
From $\eta$ we define another parameter 
$\gamma\geq 0$ such that
$\gamma=\eta/\sqrt{1-\eta^2}.$ 
The abstract counterpart of the rolling equation of Proposition \ref{prop:rolling_eq_U} is
\begin{equation}\label{eq:abstract_rolling_equation}\frac{D^{\mathbb{S}}U}{dt}=-\frac{r}{1+\gamma^2}\left(U \mathbb{S}_xU\nu(x)\right)\wedge \nu(x),\end{equation}
subject to the constraint equation \eqref{eq:abstract_constraint}. Note that the   submanifold $P$ is not defined.

\subsubsection{The equivalence between spin-geodesic and rolling systems}
The  next theorem shows that  the general spin-geodesic system  and the abstract rolling system  are two representations of the same system.

 \begin{theorem}\label{theor:abstract_rolling_equations}
 Let $(M,\langle\cdot,\cdot\rangle, B)$ be a spin-geodesic system, let $0< \eta<1$, and put
 \[\mathbb{S}:=\eta^{-1} B, \ \ \ \eta=\frac{\gamma}{\sqrt{1+\gamma^2}}.\]
 On $E=TM\oplus\, \mathbb{R} \nu$ define the metric connection  $D^{\mathbb{S}}$ by the above \eqref{eq:definition_DS}.
 Let $(x(t), v(t), W(t))$  satisfy the spin-geodesic equations \eqref{eq:spin_geodesic_system}.
Extend $W$ to $E$ by setting $W\nu=0$ and define
\begin{equation}\label{eq:U_from_W}
U=\frac1r\left(\eta^{-1}W +\nu\wedge v\right)\in \mathfrak{so}(E).
\end{equation}
Then
\begin{equation}\label{eq:constraint_W}
v=rU\nu, \ \ \ W=r\eta \Pi U\Pi,
\end{equation}
and $U$ satisfies
\begin{equation}\label{eq:abstract_rolling_theorem}
\frac{D^{\mathbb{S}}U}{dt}=-\frac{r}{1+\gamma^2}\left(U\mathbb{S} U\nu\right)\wedge\nu.
\end{equation}
Consequently, for each initial oriented orthonormal frame $\sigma_0:\mathbb{R}^{d+1}\rightarrow E_{x(0)}$, the equation
\begin{equation}\label{eq:D_sigma}
\frac{D^{\mathbb{S}}\sigma}{dt}=U\sigma, \ \ \ \sigma(0)=\sigma_0,
\end{equation}
has a unique solution, and $(x,\sigma)$ is an abstract no-slip rolling motion. Conversely, if $(x,\sigma)$ satisfies  the no-slip condition in \eqref{eq:constraint_W} in addition to \eqref{eq:abstract_rolling_theorem} and \eqref{eq:D_sigma}, 
and $W=r\eta \Pi U\Pi$,
 then $(x, v, W)$ satisfies the spin-geodesic equations \eqref{eq:spin_geodesic_system}.
Furthermore,
if $D^{\mathbb{S}}$ is flat and $M$ is simply connected, there is an isometric immersion $f:M\rightarrow \mathbb{R}^{d+1}$ whose shape
operator, with respect to the normal corresponding to $\nu$, is $\mathbb{S}$. Under a $D^{\mathbb{S}}$-parallel trivialization of $E$,
the abstract rolling equations become the ordinary Euclidean rolling equations of Proposition \ref{prop:rolling_eq_U}.
 \end{theorem}
\begin{proof}
Since $W\nu=0$ and $W(TM)\subseteq TM$, equation \eqref{eq:U_from_W} gives
\[U\nu =\frac1r(\nu\wedge v)\nu =\frac{v}{r}.\]
Hence $v=rU\nu$.  Likewise
$\Pi U\Pi =(r\eta)^{-1}W$. Thus both relations in \eqref{eq:constraint_W} hold.  
We have now to differentiate $U$ using $D^{\mathbb{S}}$. Towards this goal, we first prove the following two identities:
extending  $W$ to a skew-symmetric endomorphism of $E$ with $W\nu=0$ we have, along $x(t)$,
\begin{equation}\label{eq:DS_W_nabla_W}
\frac{D^\mathbb{S} W}{dt} = \frac{\nabla  W}{dt} +\nu\wedge W\mathbb{S} v, \ \ \ \frac{D^\mathbb{S}}{dt}(\nu\wedge v) = -(\mathbb{S} v)\wedge v +\nu\wedge\frac{\nabla v}{dt}.
\end{equation}
Let us verify first the identity on the left. Let $Y$ be a vector field on $M$. Then,  
\begin{equation*}
\begin{aligned}
\left(\frac{D^{\mathbb{S}}W}{dt}\right)Y&=   \frac{D^{\mathbb{S}}}{dt}(WY) - W\frac{D^{\mathbb{S}}Y}{dt}\\
&= \frac{\nabla}{dt}(WY) +\langle v, \mathbb{S}WY\rangle\nu  - W\frac{\nabla Y}{dt}\\
&=  \left(\frac{\nabla}{dt}W\right)Y -\langle W\mathbb{S}v,Y\rangle\nu \\
&=  \left(\frac{\nabla}{dt}W+ \nu\wedge W\mathbb{S}v\right)Y.  
\end{aligned}
\end{equation*}
Also note that 
\[\frac{D^\mathbb{S} W}{dt}\nu = \frac{D^\mathbb{S}}{dt}(W\nu) -W\frac{D^\mathbb{S} \nu}{dt} =W\mathbb{S}v=\left(\nu\wedge W\mathbb{S}v\right)\nu, \ \ \  \frac{\nabla W}{dt}\nu = \frac{\nabla W\nu}{dt} -W\frac{\nabla \nu}{dt} =0 \]
since  $\frac{\nabla \nu}{dt} =0$. 
It follows that the first identity in \eqref{eq:DS_W_nabla_W} holds when applied  to $\nu$ and to all vector fields $Y$ tangent to $M$, hence it is true in general. 
For the second identity, first note that as $D^{\mathbb{S}}$ is metric, it satisfies the product rule for the wedge product of vectors. The first term on the right-hand side of the identity follows from 
$D^{\mathbb{S}}_v \nu= -\mathbb{S}v$, and the second term follows from 
\[\frac{D^{\mathbb{S}}v}{dt}= \frac{\nabla v}{dt} +\langle v, \mathbb{S} v\rangle \nu\]
and the fact that the normal term disappears when wedged with $\nu$. 

We can now differentiate $U$. From \eqref{eq:U_from_W} and \eqref{eq:DS_W_nabla_W},
\begin{equation}\label{eq:derivative_U_1}
r\frac{D^{\mathbb{S}}U}{dt}= \eta^{-1}\frac{\nabla W}{dt} +\eta^{-1} \nu\wedge W\mathbb{S}v-(\mathbb{S}v)\wedge v +\nu\wedge \frac{\nabla v}{dt}.
\end{equation}
Recalling that $B=\eta \mathbb{S}$, we now substitute the spin-geodesic equations \eqref{eq:spin_geodesic_system} into \eqref{eq:derivative_U_1}
and obtain
\begin{equation}\label{eq:derivative_U_2}\frac{D^{\mathbb{S}}U}{dt}=\frac1r\left(\eta^{-1}-\eta\right) \nu\wedge W\mathbb{S} v = \frac{1-\eta^2}{r\eta} \nu\wedge W\mathbb{S} v.\end{equation}
Using $U\nu = v/r$,
\[U\mathbb{S} U\nu= \frac1r U\mathbb{S}v.\]
Furthermore, 
\[U\mathbb{S} v = \frac1r \left[\eta^{-1} W\mathbb{S} v +(\nu\wedge v)\mathbb{S} v\right]=\frac1r \left[\eta^{-1} W\mathbb{S} v -\langle v, \mathbb{S} v\rangle \nu\right].\]
Thus 
\[
(U\mathbb{S} U \nu)\wedge\nu = -\frac{1}{r^2\eta}\nu \wedge W\mathbb{S}v.
\]
Therefore, using $1/(1+\gamma^2)=1-\eta^2$, we obtain
\[-\frac{r}{1+\gamma^2} (U\mathbb{S}U \nu)\wedge \nu  = \frac{1}{r\eta(1+\gamma^2)}\nu\wedge W\mathbb{S} v= \frac{1-\eta^2}{r\eta}\nu\wedge W\mathbb{S} v.\]
Note that this is precisely the right-hand side of \eqref{eq:derivative_U_2}. Consequently, we obtain \eqref{eq:abstract_rolling_theorem}.

For the converse, suppose  \eqref{eq:abstract_rolling_theorem}
and \eqref{eq:D_sigma} hold. By the orthogonal decomposition of a skew endomorphism \eqref{eq:ortho_dec_skew} we have
\begin{equation}\label{eq:skew_orthogonal_U_W}
U=\frac1r \left(\eta^{-1} W + \nu \wedge v\right).
\end{equation}
Without using the spin-geodesic equations, \eqref{eq:derivative_U_1} can be grouped as
\begin{equation}\label{eq:regroup_DU}
r\frac{D^{\mathbb{S}}U}{dt}=\left[ \eta^{-1}\frac{\nabla W}{dt}-(\mathbb{S}v)\wedge v \right]+\nu\wedge\left[\eta^{-1}W\mathbb{S}v + \frac{\nabla v}{dt}\right].
\end{equation}
The first bracket expression is ``tangent-tangent'', while the second is ``normal-tangent.'' Equation \eqref{eq:abstract_rolling_theorem}
 is purely ``normal-tangent.'' Hence the first bracket expression must vanish:
 \[\frac{\nabla W}{dt}=\eta(\mathbb{S}v)\wedge v=(B v)\wedge v.\]
 Using \eqref{eq:abstract_rolling_theorem} and $1/(1+\gamma^2)=1-\eta^2$, the ``normal-tangent'' part gives
 \[\frac{\nabla v}{dt} =-\eta W\mathbb{S}v = -WBv.\]
 This concludes the equivalence between the two representations of the system. 
 
 For the second part, let us denote by $R^\mathbb{S}$ the curvature tensor of the connection $D^{\mathbb{S}}$ and by $R$ the curvature of $\nabla$.  A standard calculation in Riemannian geometry gives for vector fields $X, Y, Z$ on $M$,
 \begin{equation}\label{eq:curvature_S_1}
 R^\mathbb{S}(X,Y)Z=R(X,Y)Z+ \left(\mathbb{S}X \wedge \mathbb{S}Y\right) Z+\left(\langle Y, (\nabla_X\mathbb{S})Z\rangle -\langle X, (\nabla_Y\mathbb{S})Z\rangle\right)\nu
 \end{equation}
 and 
 \begin{equation}\label{eq:curvature_S_2}
 R^\mathbb{S}(X,Y)\nu = \left(\nabla_Y\mathbb{S}\right) X -\left(\nabla_X\mathbb{S}\right)Y.
 \end{equation}
 Observe that $\nabla_X\mathbb{S}$ is symmetric:
\begin{align*}
\langle (\nabla_X\mathbb{S})Y, Z\rangle &= \langle \nabla_X(\mathbb{S}Y),Z\rangle -\langle \mathbb{S}\nabla_XY, Z\rangle\\
&=X\langle \mathbb{S}Y,Z\rangle -\langle \mathbb{S}Y, \nabla_XZ\rangle -\langle\nabla_XY,  \mathbb{S}Z\rangle\\
&=X\langle Y,\mathbb{S}Z\rangle -\langle Y, \mathbb{S}\nabla_XZ\rangle -\langle\nabla_XY,  \mathbb{S}Z\rangle\\
&=\langle Y,\nabla_X(\mathbb{S}Z)\rangle -\langle Y, \mathbb{S}\nabla_XZ\rangle\\
&=\langle Y,  (\nabla_X\mathbb{S}) Z\rangle. 
\end{align*}
 Thus identity \eqref{eq:curvature_S_1} is equivalent to
  \begin{equation}\label{eq:curvature_S_1_bis}
 R^\mathbb{S}(X,Y)Z=R(X,Y)Z+ 
 \left(\mathbb{S}X \wedge \mathbb{S}Y\right) Z-
\langle R^\mathbb{S}(X,Y)\nu,Z \rangle \nu.
 \end{equation}
 Therefore $D^{\mathbb{S}}$ is flat exactly when the following equations hold:
 \begin{equation}\label{eq:Codazzi_Gauss}
 \left(\nabla_Y\mathbb{S}\right) X =\left(\nabla_X\mathbb{S}\right)Y, \ \ \ R(X,Y)Z+ 
 \left(\mathbb{S}X \wedge \mathbb{S}Y\right) Z=0.
 \end{equation}
 These are precisely the Codazzi and Gauss equations, respectively (with the sign convention $\mathbb{S}X=-D^{\mathbb{S}}_X\nu$.)
 The conclusion that a simply connected $M$ admits an isometric immersion as  a hypersurface in $\mathbb{R}^{d+1}$ now follows from the fundamental theorem of hypersurfaces (Bonnet's theorem).
\end{proof}

It is worth pointing out the following alternative proof of the last part of the previous theorem that bypasses the use of Bonnet's theorem and makes the conclusion  more transparent.
\begin{proposition}
Flatness of the connection $D^{\mathbb{S}}$ implies that the abstract ambient geometry, for a connected and simply connected $M$, develops into Euclidean space in which $M$ has an isometric immersion as a  hypersurface with shape operator $\mathbb{S}$.
\end{proposition}
\begin{proof}
Assume that $D^\mathbb{S}$ is flat and that $M$ is connected and simply connected.  Since $D^{\mathbb{S}}$ is a metric connection, there exists a smooth global parallel orthogonal  trivialization 
\[\Phi:E\rightarrow M\times \mathbb{R}^{d+1},\]
with $\Phi_x=\Phi(x,\cdot): E_x\rightarrow \mathbb{R}^{d+1}$ being a linear isometry.
Define an $\mathbb{R}^{d+1}$-valued one-form on $M$ as
\[\alpha_x(u):=\Phi_x(u), \ \ \ u\in T_xM\subseteq E_x.\]
Since $\Phi$ is $D^{\mathbb{S}}$-parallel, 
\begin{align*}
d\alpha(X,Y) &=X\alpha(Y)-Y\alpha(X) -\alpha([X,Y])\\
&= \Phi\left(D^{\mathbb{S}}_XY-D^{\mathbb{S}}_YX-[X,Y]\right)\\
&=\Phi\left(\nabla_XY-\nabla_YX-[X,Y]+(\langle X,\mathbb{S}Y\rangle - \langle Y,\mathbb{S}X\rangle)\nu\right)\\
&= 0,
\end{align*}
where we have used that $\nabla$ is torsion-free and that $\mathbb{S}$ is symmetric. Thus $\alpha$ is a closed form, hence exact, as $M$ is simply connected. Therefore, $\alpha=df$ for a smooth function 
\[f:M\rightarrow \mathbb{R}^{d+1}.\]
But 
\[df_x(u)=\Phi_x(u),\]
which implies that $f$ is an isometric immersion. Now set $\mathbf{n}(x):=\Phi_x(\nu)$. Then $\mathbf{n}$ is a unit normal vector to $f(M)$ and
\[d\mathbf{n}(u) = \Phi_x\left(D^{\mathbb{S}}_u \nu\right)=-\Phi_x\left(\mathbb{S}_xu\right)=-df_x(\mathbb{S}_xu).\]
But this precisely means that $\mathbb{S}$ is the shape operator of the image of $M$ under $f$. 
\end{proof}
   
Further note that under the bundle trivializing map $\Phi$ defined in the previous proof,  if 
\[A(t)=\Phi_{x(t)}\circ \sigma(t) \in \text{SO}(d+1),\]
where $\sigma(t)$ is a section of the orthonormal frame bundle over the differentiable curve $x(t)\in M$,
then, since $\Phi$ is parallel,
\[\dot{A}A^{-1} = \Phi U \Phi^{-1}\in \mathfrak{so}(d+1).\]
Thus $\frac{D^\mathbb{S}U}{dt}$ becomes the ordinary derivative  of the Euclidean angular velocity, $\nu$ becomes the Euclidean unit normal $\mathbf{n}$, and \eqref{eq:abstract_rolling_theorem}
and \eqref{eq:D_sigma} become the concrete rolling equations of Proposition \ref{prop:rolling_eq_U}.

 \subsubsection{Forced spin-geodesic systems}
 The notation and hypotheses of Theorem \ref{theor:abstract_rolling_equations} continue to hold, so that $B=\eta\mathbb{S}$, $\eta^2=\frac{\gamma^2}{1+\gamma^2}$, and the vector bundle $E=TM\oplus\, \mathbb{R}\nu$ is equipped with the metric connection $D^\mathbb{S}$.   Given a smooth curve $x(t)\in M$, let \[f(t)\in T_{x(t)}M,  \ \ \ \tau(t)\in \mathfrak{so}(T_{x(t)}M)\]
 be, respectively, an external translational force and spin torque. Then the \emph{forced spin-geodesic equations} are defined as
 \begin{equation}\label{eq:FSG_equation}
\frac{\nabla v}{dt}= -W B_x v+f, \ \ \ \frac{\nabla W}{dt}= \left(B_x v\right)\wedge v +\tau, \ \ \ \dot{x}=v.
 \end{equation}
 Under the correspondence 
 \[U=\frac1r\left(\eta^{-1} W +\nu \wedge v\right),\]
 this system is equivalent (in the sense of Theorem \ref{theor:abstract_rolling_equations}) to the no-slip constraint
 $v=rU\nu$ and the forced abstract rolling equation 
 \begin{equation}\label{eq:forced_abstract_rolling}
 \frac{D^{\mathbb{S}}U}{dt}=-\frac{r}{1+\gamma^2}\left(U\mathbb{S} U\nu\right)\wedge\nu +\frac1{r\eta} \tau +\frac1r\nu\wedge f.
 \end{equation}
 Here, $f$ and $\tau$ are regarded, respectively, as a tangential vector in $E$ (translational force) and a skew endomorphism of $E$ annihilating $\nu$ (spin torque).
 If the force and torque are induced by an \emph{abstract wrench} 
 $(\psi,\Psi)$, which is a section over $x(t)$ of $E\oplus \mathfrak{so}(E)$ obtained via the correspondence  
 \begin{equation}\label{eq:ft_wrench_correspondense}
 f=r\eta^2\Psi \nu +\left(1-\eta^2\right)\Pi\psi, \ \ \ \tau =r\eta\Pi \Psi \Pi,
 \end{equation}
 then \eqref{eq:forced_abstract_rolling} becomes
 \begin{equation}
  \frac{D^{\mathbb{S}}U}{dt}=-\frac{r}{1+\gamma^2}\left(U\mathbb{S} U\nu\right)\wedge\nu
 +\Pi\Psi\Pi +\eta^2 \nu\wedge (\Psi \nu)+\frac{1-\eta^2}{r} \nu\wedge\Pi \psi.
 \end{equation}
 Equivalently, defining the nonholonomically projected wrench
 \begin{equation}\label{eq:NH_projected_wrench}
 \mathcal{W}_\eta(\psi, \Psi)= \Pi \Psi \Pi +\eta^2 \nu\wedge (\Psi\nu) +\frac{1-\eta^2}{r} \nu\wedge\Pi \psi,
 \end{equation}
 the forced rolling equation becomes simply
 \begin{equation}\label{eq:rolling_with_Wrench_map}
  \frac{D^{\mathbb{S}}U}{dt}=-\frac{r}{1+\gamma^2}\left(U\mathbb{S} U\nu\right)\wedge\nu + \mathcal{W}_\eta(\psi,\Psi).
 \end{equation}
 If $D^\mathbb{S}$ is flat and $M$ is simply connected, the Euclidean realization supplied by Theorem \ref{theor:abstract_rolling_equations}
  carries \eqref{eq:ft_wrench_correspondense}-\eqref{eq:rolling_with_Wrench_map} into the ordinary rolling equations with wrench forces.

The proof of these claims is essentially immediate from the calculation already used in the proof of Theorem \ref{theor:abstract_rolling_equations} and will be omitted. Note that the pair \[(f,\tau)\in TM\oplus \mathfrak{so}(TM)\] is the genuinely intrinsic force on the spin-geodesic/Atiyah algebroid description, whereas \[(\psi,\Psi)\in E\oplus \mathfrak{so}(E)\] is the ambient \emph{wrench}. Equation \eqref{eq:ft_wrench_correspondense}
is then the nonholonomic  projection
\[(\psi,\Psi)\mapsto (f,\tau),\]
and \eqref{eq:forced_abstract_rolling} shows that the same projected force appears in the abstract rolling description in the form
\[F = \frac1{r\eta} \tau +\frac1r\nu\wedge f\in \mathfrak{so}(E).\]

\section{A Liouville theorem for spin-geodesic flows}\label{sec:Liouville_theorem}
Spin-geodesic systems are not in general Hamiltonian. Nevertheless, their flow preserves the canonical volume form
derived from the symplectic structure of a mechanical phase space to be introduced shortly. The system is conservative and
the induced volume form on energy level sets is also invariant under the restricted flow.

\subsection{Atiyah groupoid, algebroid, and   the orientation bundle}\label{sec:AG_AA_N}
Before turning to the Liouville theorem, we describe the several closely related spaces on which the spin-geodesic dynamics
may be represented. Let
\[p:\mathcal{F}=\text{SO}(M)\rightarrow M\] 
be the positively oriented orthonormal frame bundle of the oriented $d$-dimensional Riemannian manifold $M$, with structure group $G=\text{SO}(d)$. The tangent bundle $T\mathcal{F}$ is the unreduced velocity phase space. Reduction by the lifted right $G$-action gives the \emph{Atiyah algebroid}:
\[\mathcal{A}:=T\mathcal{F}/G\rightarrow M.\]
The Levi-Civita connection determines the familiar splitting
\begin{equation}\label{eq:splitting_Atiyah_bundle}\mathcal{A}\cong TM \oplus\mathfrak{so}(TM),\end{equation}
under which a state of the spin-geodesic system is written $(x,v,W)$, with $v\in T_xM$ and $W\in \mathfrak{so}(T_xM)$.
This is the representation in which the equations of motion take their simplest form. 
 The finite counterpart of $\mathcal{A}$ is the \emph{Atiyah} or \emph{gauge groupoid}
 \[\mathcal{G}:=(\mathcal{F}\times \mathcal{F})/G\rightrightarrows M,\]
 where $G$ acts diagonally on $\mathcal{F}\times \mathcal{F}.$ If $\sigma_1\in \mathcal{F}_{y}$ and  $\sigma_2\in \mathcal{F}_{x}$,
 the class $[\sigma_1,\sigma_2]$ represents the orientation-preserving  linear isometry 
 \[\sigma_1\circ \sigma_2^{-1}:T_xM\rightarrow T_yM.\]
The source and target maps are therefore
\[s([\sigma_1,\sigma_2])=x, \ \ \  t([\sigma_1, \sigma_2]) =y,\]
and multiplication is composition of the corresponding linear isometries. Thus $\mathcal{G}$ records finite displacement together with finite relative rotation, while its Lie algebroid $\mathcal{A}=\text{Lie}(\mathcal{G})$ records the corresponding
infinitesimal  translation and rotation velocities.  This is the finite/infinitesimal interpretation already implicit  in the description of motions by a curve
$x(t)$  together with relative rotations between tangent spaces. 

The isotropy group of $\mathcal{G}$ at $x$ is
\[\mathcal{G}_x^x=\text{SO}(T_xM).\]
Consequently, the disjoint  union 
\[N:=\text{SO}(TM)=\coprod_{x\in M}\text{SO}(T_xM)\]
is naturally the \emph{isotropy subgroupoid} of $\mathcal{G}$. At the infinitesimal level, its Lie algebra bundle is $\mathfrak{so}(TM)$, which is precisely the kernel of the anchor of $\mathcal{A}$. Thus there are parallel exact sequences
\[1\longrightarrow \text{SO}(TM)\longrightarrow \mathcal{G}\overset{(t,s)}{\longrightarrow} M\times M\longrightarrow 1\]
and 
\[0\longrightarrow \mathfrak{so}(TM)\longrightarrow \mathcal{A}\overset{\rho}{\longrightarrow} TM\longrightarrow 0.\]
The second is the Atiyah exact sequence.  Under the splitting \eqref{eq:splitting_Atiyah_bundle}, 
\[\rho(v,W)= v.\]
Sections of $\mathcal{A}$ may be identified with $\text{SO}(d)$-invariant vector fields on $\mathcal{F}$, and their Lie bracket is induced
by the ordinary vector-field bracket on $\mathcal{F}$. In the Levi-Civita splitting, and with the curvature convention
$R(X,Y)=\nabla_X\nabla_Y-\nabla_Y\nabla_X-\nabla_{[X,Y]}$, 
it is
\[\left[(X_1,W_1), (X_2,W_2)\right] =\left([X_1, X_2], \nabla_{X_1}W_2 -\nabla_{X_2}W_1 +[W_1, W_2]-R(X_1, X_2)\right).\]
Thus the Levi-Civita connection splits the Atiyah sequence as vector bundles, but not in general as Lie algebroids; its curvature measures precisely the failure of the splitting to preserve brackets.

The bundle $N=\text{SO}(TM)$ gives a useful orientation-resolved description of the same dynamics. Let
\[\pi:N\rightarrow M.\]
The Levi-Civita connection determines horizontal and vertical subspaces in $TN$, and hence a canonical 
vector-bundle isomorphism
\begin{equation}\label{eq:multiple_isomorphisms}TN\cong\pi^*\mathcal{A}\cong \pi^*\left(TM\oplus \mathfrak{so}(TM)\right).\end{equation}
Thus, at $S\in \text{SO}(T_xM)$,
\[T_SN\cong T_xM\oplus \mathfrak{so}(T_xM),\]
and a velocity $\xi \in T_SN$ is represented by its translational and rotational components $(v, W)$. This is
exactly the horizontal-vertical decomposition developed below. 

It is useful to distinguish the roles of $\mathcal{A}$ and $TN$. The algebroid $\mathcal{A}$ is the reduced
velocity phase space and contains precisely the variables occurring in the spin-geodesic equations, $(x,v,W)$.
The manifold $N$, on the other hand, retains the relative orientation $S$, and $TN$ is the corresponding orientation-resolved
phase space. Given an $\mathcal{A}$-trajectory $(x(t), v(t), W(t))$, the equation
\[\frac{\nabla S}{dt}=SW\]
reconstructs the orientation, and hence lifts the trajectory from $\mathcal{A}$ to $TN$. This is the relation
exploited later when the spin-geodesic equations are written as a second-order system on $N$. 

The two descriptions also carry compatible canonical  volume structures. The kinetic metric on 
\[\mathcal{A}_x=T_xM\oplus \mathfrak{so}(T_xM)\]
determines a fiber volume, and together with the Riemannian volume on $M$ gives a canonical volume form
\[\Omega_{\mathcal{A}} =\text{vol}_M \wedge \text{vol}_{\mathcal{A}/M}.\]
This volume also has a natural interpretation in the modular theory of Lie algebroids. Recall that the modular
line of $\mathcal{A}$ is
\[Q_{\mathcal{A}}=\text{det}\mathcal{A}\otimes \text{det}\,T^*M.\]
From the Atiyah exact sequence
\[0\rightarrow \mathfrak{so}(TM)\rightarrow \mathcal{A}\rightarrow TM\rightarrow 0\]
one obtains
\[Q_{\mathcal{A}}\cong \text{det }\mathfrak{so}(TM).\]
The trace inner product on $\mathfrak{so}(TM)$ determines an invariant unit determinant element: horizontal transport preserves this inner product, while the infinitesimal vertical action is the adjoint action of $\mathfrak{so}(d)$, whose trace is zero. Hence $\mathcal{A}$ has vanishing
modular class and is unimodular in the sense of  \cite{ELW1999}; see also the discussion in \cite{Mar2010} of invariant volumes on Lie algebroids. Under the kinetic  metric identification $\mathcal{A}\cong\mathcal{A}^*$, the corresponding density is represented by the volume form $\Omega_{\mathcal{A}}$ used here.

On $N$, equip the fibers $\text{SO}(T_xM)$ with the bi-invariant metric induced by
\[\langle W_1, W_2\rangle =\frac12 \text{Tr}\left(W_1W_2^\intercal\right)\]
and $N$ with the resulting Kaluza-Klein metric. The corresponding metric identifies $TN$ with $T^*N$, and therefore gives $TN$
 its canonical symplectic form and Liouville volume $\Omega_{TN}$. The fiberwise isomorphism \eqref{eq:multiple_isomorphisms} 
 shows that these two canonical volumes are related by 
  \begin{equation}\label{eq:volume_correspondence}
\Omega_{TN}=\nu_{\text{Haar}}\wedge \varpi^*\Omega_\mathcal{A},
 \end{equation}
where 
\[\varpi:TN\cong N\times_M\mathcal{A}\rightarrow \mathcal{A}, \ \ \ (S;v,W)\rightarrow (x; v, W),\]
and $\nu_{\text{Haar}}$ is the Riemannian Haar volume of the fibers $\text{SO}(T_xM).$ 
More precisely, $\nu_{\text{Haar}}$ is naturally a relative volume form along the fibers of $\varpi$. Indeed, the exact sequence
\[0\longrightarrow V\varpi \longrightarrow T(TN)\overset{d\varpi}{\longrightarrow} \varpi^*T\mathcal{A}\longrightarrow 0,\]
where $V{\varpi}$ is the vertical subbundle,
induces a canonical determinant-line isomorphism
\[\text{det}T^*(TN)\cong \text{det}(V{\varpi})^* \otimes \varpi^*\left(\text{det}\, T^*\mathcal{A}\right).\]
Thus the fiberwise Haar volume  and the volume form $\Omega_{\mathcal{A}}$
determine a top-degree form on $TN$. With the horizontal splitting  induced by the Levi-Civita connection, this determinant-line product is
represented by the wedge notation of \eqref{eq:volume_correspondence}. For the  Kaluza-Klein metric used here, this is precisely the symplectic volume on $TN$, with compatible choices of orientation. 
We further have
\begin{equation}\label{eq:vol_TN_and_vol_A}\varpi_{!}\Omega_{TN}=\text{Vol}(\text{SO}(d))\Omega_{\mathcal{A}}.\end{equation}
The passage from $TN$ to $\mathcal{A}$ simply forgets the reconstructed orientation.

The spin-geodesic flow on $TN$ is the reconstruction lift of the flow on $\mathcal{A}$, and its motion in the
$\text{SO}(d)$-fibers preserves Haar volume. Consequently, the Liouville theorem proved below on $TN$ has an equivalent
reduced formulation on $\mathcal{A}$: the spin-geodesic flow preserves $\Omega_\mathcal{A}$.

The same observation applies to constant-energy manifolds. The kinetic-energy function on $TN$ is the pullback under $\varpi$ of 
\[H_\mathcal{A}(x,v,W)=\frac12\left(|v|^2+\|W\|^2\right),\]
so 
\begin{equation}\label{eq:Hamiltonian_correspondence}
H_{TN} = H_\mathcal{A}\circ \varpi.
\end{equation}
Hence each energy shell in $TN$ is the $\text{SO}(d)$-bundle over the corresponding energy shell in $\mathcal{A}$.
Consequently, the same disintegration applies to every positive regular energy hypersurface: its induced
measure on $TN$ is the Haar extension, up to the same constant normalization, of the corresponding microcanonical measure on $\mathcal{A}$.

\subsection{The geometry of  $N=\text{SO}(TM)$}
Let $(M,\langle\cdot,\cdot\rangle)$ be an oriented Riemannian manifold of dimension $d$.  Define the fiber bundle
\[\pi:N=\text{SO}(TM)=\coprod_{x\in M}\text{SO}(T_xM)\rightarrow M.\] The fiber above $x\in M$ is 
 the rotation group $N_x:=\pi^{-1}(x)=\text{SO}(T_xM)$.
Let $\mathcal{V}$ denote the 
 vertical subbundle   of $TN$:
   \[\mathcal{V}_S:=\text{ker}\left(d\pi_S:T_SM\rightarrow T_{\pi(S)}M\right),\]
for each $S\in N$.
If $\xi\in \mathcal{V}_S$,  there is  $W$ in the Lie algebra $\mathfrak{so}(T_xM)$ such that
\[\xi=SW :=\left.\frac{d}{dt}\right|_{t=0} S\exp(tW).\]
If $S(t)$ is a differentiable curve in $N$ representing $\xi$, then $W=S^{-1}\dot{S}$ where $\dot{S}=S'(0)$.
We introduce on each fiber $N_x$ the   Riemannian metric 
\begin{equation}\label{eq:trace_inner_product}\langle SW_1, SW_2\rangle_S:=\frac12 \text{Tr}\left(W_1W_2^\intercal\right)\end{equation}
Here $\text{Tr}(\cdot)$ is the trace of linear maps on $T_xM$. The trace metric is a  bi-invariant Riemannian metric on each group fiber of $N$.
We also define at each $S\in N$ the {\em connection map} $K_S:T_SN\rightarrow \mathfrak{so}(T_{\pi(S)}M)$ given by
\[K_S\xi:=S^{-1}\left(\left.\frac{\nabla^M}{dt}\right|_{t=0}S(t)\right),\]
where $S(t)$ is a differentiable curve representing $\xi$ and the covariant derivative along $\pi(S(t))$ is the one induced on 
tensor fields from the Levi-Civita connection on $M$. The connection map is a linear map at each $S\in N$ and it depends smoothly on $S$. 
Note that, on vertical vectors, 
\[K_S (SW)=W.\]

Define the {\em horizontal subbundle} $\mathcal{H}$ of $TN$  so that $\mathcal{H}_S$ is the kernel of $K_S$ at each $S\in N$. 
The differential $d\pi_S$ restricts to a   linear isomorphism from $\mathcal{H}_S$ to $T_{\pi(S)}M$ and every $v\in T_xM$
admits a {\em horizontal lift} $\overline{v}_S\in \mathcal{H}_S$ defined as follows. Let $\gamma(t)$ be a differentiable curve in $M$
representing $v$, and let $S(t)$ be the parallel transport of $S$ along $\gamma$. Explicitly, if $P_t:T_{\gamma(0)}M\rightarrow T_{\gamma(t)}M$ is the
parallel transport isometry, then \[S(t)=P_{t}\circ S\circ P_{-t}.\]
It follows that $\xi:=\dot{S}$ lies in the kernel of the connection map and that $\pi(S(t))=\gamma(t)$. Thus
$d\pi_S\xi=v.$ As a consequence, $TN$ splits as a direct sum of vector subbundles
\[TN=\mathcal{V}\oplus \mathcal{H}.\]
For each $S\in N$, we have the linear isomorphism
\[\xi \in T_SN\mapsto \left(d\pi_S\xi, K_S\xi\right)\in T_{\pi(S)}M\oplus  \mathfrak{so}(T_{\pi(S)}M). \]  
 Consequently, we obtain a vector bundle isomorphism
 \[TN\cong \pi^*\left(TM  \oplus    \mathfrak{so}(TM)\right)\]
 between $TN$ and the pullback under $\pi:N\rightarrow M$ of the direct sum vector bundle over $M$.
 
 We define on $N$ the Riemannian metric that makes the vertical-horizontal splitting orthogonal and $\left.d\pi_S\right|_{\mathcal{H}_S}$
 a  linear  isometry. Explicitly, if $\xi,\zeta\in T_SN$, then
 \begin{equation}\label{eq:Riemannian_metric_N}\left\langle \xi,\zeta\right\rangle_S=\left\langle K_S\xi, K_S\zeta\right\rangle_{S} + \left\langle d\pi_S\xi, d\pi_S\zeta\right\rangle_{\pi(S)}.\end{equation}
The first term on the right-hand side is the trace inner-product on $\mathfrak{so}(T_{\pi(S)}M).$ We refer to this Riemannian metric as the \emph{Kaluza-Klein} metric on $N$.

 \subsubsection{Two connections on $TN$}
 We denote by $a:\mathcal{A}\rightarrow M$ the vector bundle with fibers
 \[\mathcal{A}_x= T_xM\oplus  \mathfrak{so}(T_xM).\] 
 This is the Atiyah bundle associated to the positive orthonormal frame bundle over $M$. The Levi-Civita connection $\nabla=\nabla^M$ on $TM$ naturally induces a connection 
 on $\mathcal{A}$. Its pullback under $\pi:N\rightarrow M$ gives a connection on $TN$, which we denote by $D$. In addition, $TN$ has its own Levi-Civita connection, denoted by $\nabla^N$, as a Riemannian manifold with 
 the Kaluza-Klein metric  \eqref{eq:Riemannian_metric_N}. We wish to compare these two connections, $D$ and $\nabla^N$, on $TN$. 
 
 We use the following notation: for a basic vector field $X\in \mathfrak{X}(M)$ let $X^H\in \mathfrak{X}(N)$ denote the horizontal lift of $X$, and for
 a smooth section $W\in \Gamma(\mathfrak{so}(TM))$ denote by 
$ W^V\in \mathfrak{X}(N)$ the vertical vector field  such that $W^V_S=SW$. 
Then $D$ is characterized by the following identities:  
  \begin{equation}\label{eq:derivatives_D}D_{X^H}Y^H=\left(\nabla_X^MY\right)^H, \ \ D_{X^H}W^V =\left(\nabla_XW\right)^V, \ \  D_{W^V}Y^H=0, \ \ D_{W^V} U^V=0.\end{equation}
  \begin{proposition}
   $D$ is a metric connection on $TN$.
  \end{proposition}
  \begin{proof}
  Let us differentiate the inner product
  \[\left\langle X_1^H+W_1^V, X_2^H+W_2^V\right\rangle= \langle X_1,X_2\rangle\circ\pi +\frac12\text{Tr}\left(W_1W_2^\intercal\right)\]
  along the vector field $Y^H+U^V$, where $X_1, X_2,Y\in \mathfrak{X}(M)$ and $W_1,W_2,U\in \Gamma(\mathfrak{so}(TM))$.
  This inner product is constant on fibers and the terms coming from the $D_{U^V}$ vanish identically. So the metric property holds under
  covariant differentiation along $U^V$. Now, differentiation in $D_{Y^H}$ satisfies the metric property when applied 
  to the term $ \langle X,Y\rangle\circ\pi $ since this reduces to the metric property of $\nabla^M$ on $TM$. Thus we only need to check that
  \[Y \text{Tr}\left(W_1W_2^\intercal \right)=\text{Tr}\left(\left(\nabla^M_YW_1\right)W_2^\intercal\right) +\text{Tr}\left(W_1\left(\nabla^M_YW_2\right)^\intercal\right).  \]
  But this  reduces to the metric property of $\nabla^M$ on $TM$ due to the identity
  \[\langle X_1\wedge Y_1, X_2\wedge Y_2\rangle = \langle X_1, X_2\rangle \langle Y_1, Y_2\rangle -\langle X_1, Y_2\rangle\langle X_2, Y_1 \rangle,\]
  where $X_1, X_2, Y_1, Y_2$ are vector fields on $M$. 
  \end{proof}

 \begin{proposition}\label{prop:cal_C}
 Let $R^M$ be the curvature tensor of the Levi-Civita connection $\nabla^M$ on $M$. For $X,Y\in \mathfrak{X}(M)$ and $S\in N_x$ define $\mathcal{C}_S(X,Y)\in \mathfrak{so}(T_xM)$ by 
 \[\mathcal{C}_S(X,Y)^V:=\left[X^H,Y^H\right] - [X,Y]^H,\]
 where the Lie brackets are evaluated at $S$.
 Then, setting $x=\pi(S)$, 
 \[\mathcal{C}_S(X,Y)= R_x^M(X,Y)- S^{-1} \circ  R_x^M(X,Y)\circ S.\]
 In particular, $\mathcal{C}_{I_x}(X,Y)=0$ at the identity endomorphism $I_x$.  
 \end{proposition} 
  \begin{proof}
  First observe that if $S$ is a field of endomorphisms on $TM$, then $\nabla^M$ acts on $S$ by 
  \[(\nabla^M_XS)Y=\nabla_X^M (SY) - S\nabla^M_XY\] for all $X,Y\in \mathfrak{X}(M)$. The curvature of $\nabla^M$ extended to $\text{End}(TM)$ is shown by a straightforward calculation to satisfy
  \begin{equation}\label{eq:R_commutator_S}R^M(X,Y)S =R^M(X,Y)\circ S -S\circ R^M(X,Y)=[R^M(X,Y), S].   \end{equation}
 There is a tautological section $\mathcal{R}\in \Gamma(\pi^*\text{End}(TM))$  defined by
 \[\mathcal{R}(S)=S. \] 
 If $\xi\in T_SN$ is represented by a curve $S(t)\in \text{SO}(T_{x(t)}M)$ then (denoting $\nabla:=\nabla^{\pi^*\text{End}(TM)}$)
 \[\nabla_\xi \mathcal{R}=\left.\frac{DS(t)}{dt}\right|_{t=0}.\]
It follows that  $\xi$ is a horizontal vector precisely when 
\begin{equation}\label{eq:zero_horizontal_derivative_R}\nabla_\xi \mathcal{R}=0.\end{equation}
In other words, a  curve $S(t)$  in $N$ is horizontal precisely when the orthogonal endomorphism itself is parallel.
Equivalently, if $P_t:T_xM\rightarrow T_{x(t)}M$ is parallel transport, then \[S(t) = P_t S(0) P_t^{-1}.\] 

It will be helpful to define the $1$-form $\varrho$ on $N$ taking values in $\mathfrak{so}(TM)$ by
\[\varrho_S(\xi):=S^{-1}\nabla_\xi\mathcal{R}\in \mathfrak{so}(T_{\pi(S)}M).\]
Then \begin{equation}\label{eq:omega_horizontal_vertical}\varrho_S\left(X^H\right)=0, \ \ \ \varrho\left(W^V\right)=W.\end{equation}

Let now $X, Y\in \mathfrak{X}(M)$ and $X^H$, $Y^H$ be the respective horizontal lifts to $N$. Since these lifts are projectable vector fields,
 $\left[X^H,Y^H\right] - [X,Y]^H$ is, at each $S\in N$, a vertical vector, so there must exist an endomorphism $\mathcal{C}_S(X,Y):T_xM\rightarrow T_xM$, which is tensorial in the two vector arguments,  such that the difference of brackets is $\mathcal{C}_S(X,Y)^V$. Note that
 \[S\mathcal{C}_S(X,Y)=\mathcal{C}_S(X,Y)^V.\] We next determine $\mathcal{C}_S(X,Y)$. We use the curvature of the pulled-back
 $\text{End}(TM)$-connection $\nabla$ on the tautological section $\mathcal{R}$:
 \[R^\nabla\left(X^H,Y^H\right)\mathcal{R}=\nabla_{X^H}\nabla_{Y^H}\mathcal{R}-\nabla_{Y^H}\nabla_{X^H}\mathcal{R}-\nabla_{[X^H,Y^H]}\mathcal{R}.\]
As the first two terms vanish by (\ref{eq:zero_horizontal_derivative_R}), we have 
\[R^\nabla\left(X^H,Y^H\right)\mathcal{R}=-\nabla_{[X^H,Y^H]}\mathcal{R}.\]
 Since the pullback curvature is the base curvature, and due to (\ref{eq:R_commutator_S}),  we have at $S$ 
\[ R_S^\nabla\left(X^H,Y^H\right) \mathcal{R}= R_x(X,Y)S= R_x(X,Y)\circ S - S\circ R_x(X,Y), \]
where $x=\pi(S)$.
Therefore
\[\left(\nabla_{[X^H,Y^H]}\mathcal{R}\right)_S=  S\circ R_x(X,Y)-R_x(X,Y)\circ S.\]
Composing this expression on the left by $S^{-1}$ gives
\[\varrho_S\left(\left[X^H,Y^H\right]\right)= R_x(X,Y) - S^{-1}\circ R_x(X,Y)\circ S.\]
Since $\varrho$ vanishes when evaluated on horizontal vectors,
$\varrho_S\left(\left[X^H,Y^H\right]\right) =\mathcal{C}_S(X,Y)$ due to
 (\ref{eq:omega_horizontal_vertical}),
hence 
\[\mathcal{C}_S(X,Y) = R_x(X,Y)-S^{-1} \circ R_x(X,Y)\circ S\]
as claimed.
  \end{proof}
  
  \begin{corollary}
  If $M$ is a two-dimensional Riemannian manifold, then $\mathcal{C}=0$.
  \end{corollary}
  \begin{proof}
  This is due to the fact that every orientation-preserving $S$ commutes with the single generator $J\in \mathfrak{so}(2)$. 
  \end{proof}
  
  \begin{corollary}
  If $M$ has constant sectional curvature $\kappa$, 
  \[\mathcal{C}_S(X,Y)=-\kappa\left[X\wedge Y -\left(S^{-1}X\right)\wedge \left(S^{-1}Y\right)\right].\]
  \end{corollary}
  \begin{proof}
  The Riemannian curvature tensor of a metric with constant sectional curvature $\kappa$ has the form
  $R(X,Y) = \kappa X\wedge Y.$ The claim follows by noting that
  \[S (X\wedge Y) S^{-1} =(SX)\wedge (SY)\] for any orthogonal transformation $S$.
    \end{proof}

  \subsubsection{Torsion of the pulled-back connection}
  
 \begin{proposition}\label{prop:brackets_HV}
 Let $X\in \mathfrak{X}(M)$ and $W, W_1, W_2 \in \Gamma(\mathfrak{so}(TM))$. Then
 \[\left[X^H,W^V\right]_S= S \left(\nabla_XW\right)_x, \ \ \left[W_1^V,W_2^V \right]_S = [W_1,W_2]_x^V\]
 at  $S\in N_x$.
 \end{proposition}
 \begin{proof}
 Let   $\Phi_t$ be the flow of the horizontal lift $X^H$ of $X$ on $N$. 
  As the curve $S(t)=\Phi_t(S)$ is horizontal, we have seen that 
  \[S(t) = P_t\circ S\circ P_t^{-1}\]
  where $P_t$ is parallel transport along $x(t)=\pi(\Phi_t(S))$. Let now $W$ be a local section of $\mathfrak{so}(TM)$.
 The Lie derivative  of $W^V$ along $X^H$ at $S$ is
 \[\left[X^H, W^V\right]_S =\left. \frac{d}{dt}\right|_{t=0} \left(\Phi_{-t}\right)_*\left(W^V_{S(t)}\right)=
 \left. \frac{d}{dt}\right|_{t=0} P_{-t}   W^V_{S(t)}   P_{-t}^{-1}= \left. \frac{d}{dt}\right|_{t=0} P_t^{-1}   W^V_{S(t)}   P_{t}.\]
  At $S(t)$ the vertical field is \[W^V_{S(t)}=S(t) W_{x(t)}=P_tSP_t^{-1}W_{x(t)}.\]
 Therefore
 \[\left(\Phi_{-t}\right)_*\left(W^V_{S(t)}\right)=P_t^{-1}\left(P_tSP_t^{-1}W_{x(t)}\right)P_t=SP_t^{-1}W_{x(t)}P_t.\]
 Differentiating at $t=0$,
 \[\left[X^H,W^V\right]_S = S\left. \frac{d}{dt}\right|_{t=0} \left(P_t^{-1}W_{x(t)}P_t \right)=S \left(\nabla_XW\right)_x.\]
 The second bracket identity is a standard fact about the Lie bracket of left-invariant vector fields on a Lie group. 
 \end{proof}
 
 \begin{proposition}
 The torsion tensor  $T^D$ of the pulled-back connection $D$ on $TN$ at $S\in N$ satisfies the identities
 \[T^D\left(X^H,Y^H\right)=-\mathcal{C}_S(X,Y)^V, \ \ T^D\left(X^H,W^V\right)=0, \ \ T^D\left(W_1^V,W_2^V\right)=-[W_1, W_2]^V,\]
 where $X\in \mathfrak{X}(M)$ and $W, W_1, W_2 \in \Gamma(\mathfrak{so}(TM))$. 
 \end{proposition}
 \begin{proof}
 The first identity follows from the definition of $\mathcal{C}$ and that $D_{X^H}Y^H=\left(\nabla^M_XY\right)^H$. 
 The other two identities are a consequence of Proposition \ref{prop:brackets_HV}.
 \end{proof}
 Thus $D$ is a metric connection on $TN$ with a torsion tensor that reflects the intrinsic curvature of $M$ and the noncommutativity of the
 Lie algebra fibers through their Lie bracket.
 \subsubsection{The Levi-Civita connection on $N$}\label{sec:LC_connection_C}
 Let $\nabla^N$ denote the Levi-Civita connection on $N$ for the Riemannian metric \eqref{eq:Riemannian_metric_N}.
  Define at $S\in N_x$ the vector 
  \[\mathcal{K}_S(X,W)\in T_xM\]
  for $X\in T_xM$ and $W\in \mathfrak{so}(T_xM)$
  by \begin{equation}\label{eq:definition_K}
  \left\langle \mathcal{K}_S(X,W),Y\right\rangle=\left\langle W, \mathcal{C}_S(X,Y)\right\rangle_{\mathfrak{so}(T_xM)}.
  \end{equation}
  Recall that the inner-product on $\mathfrak{so}(T_xM)$ is given by \eqref{eq:trace_inner_product}.
  
  \begin{proposition}\label{prop:map_CC}
  Let $X, Y\in \mathfrak{X}(M)$ and $W, W_1, W_2\in \Gamma(\mathfrak{so}(TM))$. Then the following identities hold:
  \begin{equation}\label{eq:Levi_Civita_N}
  \begin{aligned}
  \nabla^N_{X^H}Y^H & = \left(\nabla^M_XY\right)^H +\frac12 \mathcal{C}_S(X,Y)^V, &  \nabla^N_{W^V}X^H &=-\frac12 \mathcal{K}_S(X,W)^H\\
  \nabla^N_{X^H} W^V&=\left(\nabla^M_XW\right)^V -\frac12 \mathcal{K}_S(X,W)^H, &  \nabla^N_{W_1^V}W_2^V&=\frac12[W_1, W_2]^V.\\
  \end{aligned}
  \end{equation}
  Writing $C:=\nabla^N-D$, it follows from Equations \eqref{eq:derivatives_D} that 
\begin{equation}\label{eq:C_one}
  C_{X^H}Y^H  = \frac12 \mathcal{C}_S(X,Y)^V, \ \   C_{W^V}Y^H =-\frac12 \mathcal{K}_S(Y,W)^H, \ \ 
 C_{X^H} W^V=-\frac12 \mathcal{K}_S(X,W)^H, 
  \end{equation}
  and 
  \begin{equation}\label{eq:C_two}
   C_{W_1^V}W_2^V=\frac12[W_1, W_2]^V=D^{\text{\em ad}}_{W_1^V}W_2^V ,
  \end{equation}
  in which $D^{\text{\em ad}}$ is the Levi-Civita connection on the rotation Lie group for the bi-invariant metric. (Here $\text{\em ad}$ stands for `{adapted.}')
  In particular, along the identity section $x\mapsto I_x$,
  \[\nabla^N_{X^H}Y^H = D_{X^H}Y^H, \ \ \nabla^N_{X^H}W^V = D_{X^H}W^V, \ \ \nabla^N_{W^V} Y^H=D_{W^V}Y^H=0\]
  while, on the group fibers,
  \[\nabla^N_{W_1^V}W_2^V = D_{W_1^V}W_2^V +\frac12[W_1, W_2]^V.\]
  If instead of the pull-back connection $D$ one defines the {\em adapted} metric connection $D^{\text{\em ad}}$ by supplementing the pullback connection with the fiber Levi-Civita connection 
  $D^{\text{\em ad}}_{W_1^V}W_2^V,$
  then $\nabla^N=D^{\text{\em ad}}$ along the identity section. 
  \end{proposition}
  \begin{proof}   
  We need to establish equations \eqref{eq:Levi_Civita_N}. The other equations are a consequence of
  the   description of $D$ previously obtained.  
  We use the Koszul formula for the Levi-Civita connection, the bracket relations in Proposition \ref{prop:brackets_HV}, the definition
  of $\mathcal{C}$ in Proposition \ref{prop:cal_C}, and of $\mathcal{K}$ in equation \eqref{eq:definition_K}, as well as
the Riemannian metric on $N$ given in \eqref{eq:Riemannian_metric_N}.
The horizontal component of $\nabla^N_{X^H}Y^H$ is easily seen to be $\nabla_X^MY$ by comparing the Koszul expressions of these two connections. To determine the vertical component note that, from the Koszul identity,  after simplification using the bracket relations and
the orthogonality of vertical and horizontal components, we obtain
\[\left\langle \nabla^N_{X^H}Y^H, W^V\right\rangle_S =\frac12\left\langle \left[X^H,Y^H\right], W^V\right\rangle_S = \frac12\left\langle\mathcal{C}_S(X,Y)^V, W^V\right\rangle.\]
This gives the above form of $ \nabla^N_{X^H}Y^H$. To obtain the expression for $\nabla^N_{W_1^V}W_2^V$,  first observe that
\begin{equation}\label{eq:W1W2Z}
\begin{aligned}\left\langle \nabla^N_{W_1^V}W_2^V, Z^H \right\rangle_S&=-\frac12\left(Z^H\left\langle W_1^V, W_2^V\right\rangle +\left\langle \left[W_1^V,Z^H\right],W_2^V\right\rangle_S
+\left\langle \left[W_2^V,Z^H\right],W_1^V\right\rangle_S\right)\\
&=-\frac12\left(Z_{\pi(S)}\langle W_1, W_2\rangle -\left\langle \nabla^M_ZW_1,W_2\right\rangle_{\pi(S)}
-\left\langle \nabla^M_ZW_2,W_1\right\rangle_{\pi(S)}\right)\\
&=0,
\end{aligned}
\end{equation}
since $\nabla^M$ is a metric connection on    $\mathfrak{so}(TM)$ with the trace metric.
As  inner products of the form $\left\langle W_1^V,W_2^V\right\rangle$ are constant along fibers, and making use of the
cyclic symmetry of the trace, we have
\[\left\langle \nabla^N_{W_1^V}W_2^V, W_3^V \right\rangle_S=-\frac12\text{Tr}\left([W_1, W_2]W_3\right)=\frac12 \left\langle \left[W_1, W_2\right]^V, W^V_3 \right\rangle_{S}.\]
From these two identities, the claimed expression for $\nabla^N_{W_1^V}W_2^V$ follows.
By the torsion-free property of $\nabla^N$, we have
\[\nabla_{X^H}^NW^V-\nabla^N_{W^V}X^H=\left[X^H,W^V\right]=\left(\nabla^M_XW\right)^V.\]
Thus once we establish the expression for $\nabla^N_{W^V}X^H$, the one for $\nabla_{X^H}^NW^V$ will immediately follow.
From \eqref{eq:W1W2Z} and the metric property of $\nabla^N$ we obtain
\[\left\langle \nabla^N_{W_1^V}X^H, W_2^V \right\rangle=0.\]
Thus $\nabla^N_{W^V}X^H$ must be horizontal. To determine its horizontal component, note:
\[\left\langle \nabla^N_{W^V}X^H, Y^H\right\rangle=\frac12\left(W_1^V\left\langle X^H,Y^H\right\rangle - \left\langle \left[X^H,Y^H\right],W^V\right\rangle\right).\]
The first inner product on the right-hand side is constant on fibers, so
\[\left\langle \nabla^N_{W^V}X^H, Y^H\right\rangle_S=-\frac12 \left\langle \left[X^H,Y^H\right],W^V\right\rangle_S=-\frac12 \left\langle \mathcal{C}_S(X,Y)^V,W^V\right\rangle,\]
and the expression for $\nabla^N_{W^V}X^H$ follows from the definition of $\mathcal{K}_S(X,W)$.
  \end{proof}
  
  \subsection{The spin-geodesic equations in $\mathcal{A}$ and $N$}
  The spin-geodesic equations are expressed most simply on the Atiyah bundle  $\mathcal{A}$. 
   Let $B$ be a symmetric field of endomorphisms on $M$, which we call the {\em spin-coupling field}. Elements of $\mathcal{A}$ are written as $(v,W)$ and elements of $M$ as $x$.
  Then
  \[\dot{x}=v, \ \ \  \frac{\nabla^M v}{dt}=-WBv, \ \ \ \frac{\nabla^M W}{dt}=(Bv)\wedge v.\]
 To these equations we may add a rotation reconstruction equation
  \[\frac{\nabla^M S}{dt}=SW,\]
  which gives the tangential rotation \[S(t):T_{x(t)}M\rightarrow T_{x(t)}M\] produced by the spin velocity $W(t)$ along $x(t)$.   The resulting
  $S(t)$ is an element of $\text{SO}(T_{x(t)}M)$. If $S(0)=I_{x(0)}$, 
  the Atiyah  groupoid arrow \[g(t):T_{x(0)}M\rightarrow T_{x(t)}M\] is obtained from $S(t)$ as $g(t)=S(t)\circ P_{t}$ where $P_t:T_{x(0)}M \rightarrow T_{x(t)}M$ is parallel transport.
    
  It is useful to also describe the motion on the extended configuration manifold $N$ that   directly incorporates the rotation $S$.
  The usefulness lies, in particular, in that   $TN$ is   a symplectic manifold whose  Liouville volume arises in the most standard fashion. (Naturally, we could equally  work on the more canonical momentum space $T^*N$.)  This allows us to compare  the  volume form on the Atiyah algebroid and the   volume form on $N$ derived from the symplectic form regarding their invariance  under the spin-geodesic flow. See the remarks made 
  in subsection \ref{sec:AG_AA_N}.
  
  Recall that $TN$ is isomorphic to the pullback of $\mathcal{A}$ under the base-point projection $\pi:N\rightarrow M$. 
  Then
\begin{equation}\label{eq:split_xi_HV}\xi:=\dot{S}=v^H +W^V, \ \ v=\dot{x}.\end{equation}
  Thus the spin-geodesic equations that use the covariant derivative $\nabla^M$ combine into the single equation using the pulled-back connection $D$:
  \begin{equation}\label{eq:sg_equation_in_D}\frac{D\xi}{dt} =\left(-WBv\right)^H +\left((Bv)\wedge v\right)^V.\end{equation}
  The reconstruction equation for $S(t)$ is no longer a separate dynamical equation once $\xi=\dot{S}$ is used. 
  
  We wish now to rewrite the second order equation of motion \eqref{eq:sg_equation_in_D} in terms of $\nabla^N$. Recall that $\nabla^N=D+C$, where $C$ (from Proposition \ref{prop:map_CC}) is derived  from $\mathcal{C}$ (introduced in Proposition \ref{prop:cal_C}).  Also recall the tensor $\mathcal{K}_S(X,W)$   introduced in definition \eqref{eq:definition_K}. Then, using equations \eqref{eq:C_one} and  \eqref{eq:C_two}, we obtain
  \[C_\xi \xi =\frac12\mathcal{C}_S(v,v)^V -\mathcal{K}_S(v,W)^H  +\frac12[W,W]^V= -\mathcal{K}_S(v,W)^H.\]
  Therefore
  \begin{equation}\label{eq:sg_equation_in_nablaN}\frac{\nabla^N\xi}{dt}= \left[-WBv-\mathcal{K}_S(v,W)\right]^H +\left[(Bv)\wedge v\right]^V,\end{equation}
  to which should be added Equation \eqref{eq:split_xi_HV}.

  When expressed in terms of the Levi-Civita connection on $N$, the spin-geodesic equations acquire the additional term $\mathcal{K}_S(v,W)^H$, which
  acts as a force of geometric origin. In particular, when the coupling field $B$ (or the parameter $\eta$) is zero, $S(t)$ is not a geodesic relative to the Kaluza-Klein metric on $N$, except  when $M$ is $2$-dimensional, in which case $\mathcal{C}_S=0$ and $\mathcal{K}_S=0$ or, in any dimension, when $M$ is flat. 
  Nevertheless, energy is still preserved, as shown in the next proposition.

   \begin{proposition}\label{prop:force_orthogonal_to_velocity}
   Let $S(t)\in N$ be a solution to the spin-geodesic equation \eqref{eq:sg_equation_in_nablaN} and write $\xi(t)=S'(t)$. Then
   \[\frac{d}{dt}\|\xi\|^2 = 2\left\langle \xi, \frac{\nabla^N \xi}{dt} \right\rangle =0,\]
in which we are using the Kaluza-Klein Riemannian metric  on $N$.
   \end{proposition} 
   \begin{proof}
   We have
   \[\left\langle \xi, \frac{\nabla^N \xi}{dt} \right\rangle =-\left\langle v, WBv  \right\rangle -\left\langle v, \mathcal{K}_S(v,W)\right\rangle +\left\langle W, (Bv)\wedge v\right\rangle.  \]
   The first two inner-products on the right-hand side come from the Riemannian metric on $M$, and the third is the trace inner-product on $\mathfrak{so}(T_xM)$. From the definition of $\wedge$ we obtain
   \[ \left\langle W, (Bv)\wedge v\right\rangle = \left\langle v, WBv  \right\rangle. \]
   And from the definition of $\mathcal{K}_S(v,W)$ we have
   \[\left\langle v, \mathcal{K}_S(v,W)\right\rangle = \left\langle W, \mathcal{C}_S(v,v)\right\rangle = 0. \]
   Therefore, $\xi$ is perpendicular to  ${\nabla^N \xi}/{dt}$ as claimed.
   \end{proof}
   It follows from the previous proposition that the total kinetic energy $\|v\|_M^2 +\|W\|_{\mathfrak{so}}^2$ is a constant of motion, 
   a result of the fact that the force $F_B:TN\rightarrow TN$, defined as
   \[F_B(S,\xi):= \left[-WBv-\mathcal{K}_S(v,W)\right]^H +\left[(Bv)\wedge v\right]^V,\]
   is orthogonal to $\xi=v^H+W^V$. 
      
   \subsubsection{Spin-geodesic spray}
   Recall that $\nabla^N$ is the Levi-Civita connection 
  for the Kaluza-Klein Riemannian metric \eqref{eq:Riemannian_metric_N}. This connection gives rise to the splitting
  \[T\mathcal{N}=\mathcal{V}^\mathcal{N}\oplus\mathcal{H}^\mathcal{N}\]
  into vertical and horizontal vectors in $T\mathcal{N}$. 
     We indicate horizontal and vertical lifts of vectors tangent to $N$ by
the superscripts $H$ and $V$. It should be clear from the context when these, rather than the lifts from $TM$ to $TN$, are meant. 
  
  The horizontal and vertical splitting of $T\mathcal{N}$ allows us to define the (Sasaki) Riemannian metric on $\mathcal{N}$
  as
  \[\left\langle X_1, X_2\right\rangle_\xi  :=\left\langle K_\xi X_1, K_\xi X_2\right\rangle_S + \left\langle d\pi_\xi X_1, d\pi_\xi X_2\right\rangle_S\]
  for $\xi\in T_SN$.

  Let us express the spin-geodesic equation as a first order equation on $\mathcal{N}=TN$.
  First observe the following general fact. 
  \begin{proposition}\label{prop:lifting_system_to_TN}
  Let $Q$ be a Riemannian manifold, $\pi:TQ\rightarrow Q$ the base-point map, and $F:TQ\rightarrow TQ$ a smooth map such that $\pi\circ F= \pi.$ Then $(x(t),v(t))$  is a solution to 
\[ v(t)=\dot{x}(t), \ \   \frac{\nabla v}{dt}=F(v)\]
  if and only if $v(t)$ is an integral curve of the vector field $X$ on $\mathcal{Q}=TQ$ 
  defined by
  \[d\pi_v X =v, \ \ K_vX=F(v),\]
  where $K_v:T_v\mathcal{Q}\rightarrow T_xQ$ is the connection map: $K_v \dot{v} =\frac{\nabla v}{dt}.$ 
  Note that \[X(v)=v^H+F(v)^V\]
  where $v^H$ indicates   (Levi-Civita connection) horizontal lift and $F^V$ is the vertical lift. 
  \end{proposition}
  \begin{proof}
  Let $v(t)$ be an integral curve of $X$ and $x(t)=\pi(v(t))$. Then 
  \[\frac{\nabla v}{dt}= K_v \dot{v} =K_v X(v)=F(v).\]
  Conversely, if $v(t)=\dot{x}(t)$ satisfies  this equation along the curve $x(t)=\pi(v(t))$, then 
  \[d\pi_v \dot{v}= \frac{d}{dt}\pi(v(t))=\dot{x}(t)=v(t), \ \ K_v\dot{v}=F(v).\]
  Therefore $\dot{v}=v^H+F(v)^V=X(v)$.
  \end{proof} 
  
  Applying this proposition to the spin-geodesic equation
  \[\xi(t)=\dot{S}(t), \ \ \frac{\nabla^N \xi}{dt} = F_B(S,\xi),\]
  we conclude that $S(t)$ is a solution if and only if   $\dot{S}(t)$ is an integral curve of the 
  vector field $X^B\in \mathfrak{X}(\mathcal{N})$ defined as
  \[X^B(\xi) =\xi^H + F_B(S,\xi)^V, \]
  with $S(t)=\pi(\xi(t))$. 
  The horizontal vector field $X^0$ on $T\mathcal{N}$ such that $d\pi_\xi X^0(\xi) =\xi$ is the geodesic spray of the Riemannian manifold $N$. 
  Its integral lines project to geodesics on $N$. Setting $Y^B:=F_B^V$, we write
  \[X^B=X^0+Y^B.\]
  Thus $X^B$ is a perturbation of the geodesic spray $X^0$ of $N$  by a vertical vector field $Y^B$ parametrized by the spin-coupling field $B$.     
  When convenient, we write $Y$ instead of $Y^B$.
  
 The following result is well-known in Riemannian geometry (see, for example, \cite{Poor81}), but  due to its importance in what follows, we provide a proof so as to make the discussion largely self-contained. 
  
  \begin{proposition}\label{prop:bracket_relations}
  Let $Q$ be a Riemannian manifold with Levi-Civita connection $\nabla$ and let $\pi:TQ\rightarrow Q$ be the tangent bundle projection. Let $R$ be the curvature tensor for $\nabla$.  For vector fields $X,Y\in \mathfrak{X}(Q)$, 
  denote their horizontal and vertical lifts to $TQ$ by $X^H, Y^H$ and $X^V, Y^V$, respectively. Then
  the following bracket relations hold  for all $x$ and $v\in \pi^{-1}(x)$:
  \begin{enumerate}
  \item  $\left[X^V,Y^V\right]_v=0$;
  \item   $\left[X^H,Y^V\right]_v=\left(\nabla_XY\right)_v^V$;
 \item  $\left[X^H,Y^H\right]_v=[X,Y]_v^H-(R(X,Y)v)_v^V$.
  \end{enumerate}
Here we use the common sign convention  $R(X,Y)=\nabla_X\nabla_Y-\nabla_Y\nabla_X-\nabla_{[X,Y]}$. 
  \end{proposition}
  \begin{proof}
  Let  $\mathcal{Q}=TQ$ and $\mathcal{Q}_x=T_xQ$. 
 Define   the tautological  section $\mathcal{V}$ of the pullback bundle $\pi^*TQ\rightarrow \mathcal{Q}$  by \[\mathcal{V}(v)=v.\]
 Denote by $\nabla^\mathcal{Q}$ the pullback connection on $\pi^*TQ$. 
  Associated with $\nabla$ is the connection map
  \[K:T\mathcal{Q}\rightarrow TQ,\]
  which can be expressed in terms of $\mathcal{V}$ as
  \[K_v\xi =\nabla^\mathcal{Q}_\xi \mathcal{V},\]
  where $\xi\in T_v\mathcal{Q}$. Thus, for $X,Y\in \mathfrak{X}(Q)$, 
  \[K\left(X^H\right)=0, \ \ \ K\left(X^V\right)=X.\]
  Indeed, along a horizontal curve $v(t)$, the vector $v(t)$ is parallel, whereas along the vertical curve $v+tX_x$,
  \[\left.\frac{\nabla}{dt}\right|_{t=0} \left(v+tX_x\right)=X_x. \]
  The integral curves of $X^V$ are the fiber translations $v\mapsto v+tX_x$. Since fiber translations commute, we have the first bracket relation:
  $\left[X^V,Y^V\right]=0.$ Moving on to the second bracket relation, let $\Phi_t$ be the local flow of $X^H$. If $x(t)$ is the integral curve of $X$ through $x$ and 
  \[P_t:T_xQ\rightarrow T_{x(t)}Q\]
  is parallel transport, then $\Phi_t(v)=P_tv$. At $P_tv$, the vertical field $Y^V$ is represented by $Y_{x(t)}$.  Pulling it back by $\Phi_{-t}$ gives the
  vertical vector represented by 
  \[P_t^{-1}Y_{x(t)}\in T_xQ.\]
  Consequently,
  \[\left[X^H,Y^V\right]_v = \left.\frac{d}{dt}\right|_{t=0} d(\Phi_{-t})_{{\Phi_t(v)}} Y^V_{\Phi_{t}(v)}=\left(\left.\frac{d}{dt}\right|_{t=0} P_t^{-1} Y_{x(t)}\right)^V =(\nabla_XY)^V.\]
  Finally, consider the horizontal-horizontal bracket relation.  Since $X^H, Y^H$ are $\pi$-related to $X,Y$,
 it follows that $\left[X^H, Y^H\right]$ is $\pi$-related to $[X,Y]$. Hence   $\left[X^H, Y^H\right]-[X,Y]^H$ is vertical. Thus there is   $Z\in T_xQ$
 such that 
 \begin{equation}\label{eq:difference_H_Z}\left[X^H, Y^H\right]_v=[X,Y]_v^H+Z_v^V.\end{equation}
 It remains to determine $Z$. Let us apply the curvature $R^\mathcal{Q}$  of the pullback connection $\nabla^{\mathcal{Q}}$ to the tautological section $\mathcal{V}$:
 \[R^\mathcal{Q}\left(X^H,Y^H\right)\mathcal{V}=\nabla^{\mathcal{Q}}_{X^H}\nabla^{\mathcal{Q}}_{Y^{H}}\mathcal{V}-\nabla^{\mathcal{Q}}_{Y^H}\nabla^{\mathcal{Q}}_{X^{H}}\mathcal{V}-\nabla^{\mathcal{Q}} _{\left[X^H,Y^H\right]}\mathcal{V}.\]
 Since horizontal vectors lie in the kernel of the connection map for $\nabla^{\mathcal{Q}}$, the covariant derivative  of $\mathcal{V}$
   for this connection is $0$ along horizontal vectors.  Thus
 \begin{equation}\label{eq:curvature_HHV}
 R^{\mathcal{Q}}\left(X^H, Y^H\right)\mathcal{V} = -K\left(\left[X^H, Y^H\right]\right).
 \end{equation}
 On the other hand, the curvature of a pullback connection is the pullback of the original curvature. Therefore, at $v\in T_xQ$,
 \begin{equation}\label{eq:relation_between_curvatures}
 R_v^{\mathcal{Q}}\left(X^H,Y^H\right)\mathcal{V} = R_x(X,Y)v.
 \end{equation}
 Combining \eqref{eq:curvature_HHV} and \eqref{eq:relation_between_curvatures}, 
 \[K_v\left(\left[X^H,Y^H\right]\right)=-R_x(X,Y)v.\]
 Since $K\left([X,Y]^H\right)=0$, \eqref{eq:difference_H_Z} implies 
 \[Z=-R(X,Y)v.\]
 Therefore,
  \[\left[X^H,Y^H\right]_v=[X,Y]_v^H-(R(X,Y)v)_v^V.\]
  This completes the proof.
  \end{proof}
  
  \subsection{Volume invariance}\label{sec:volume_invariance_final}
  We are interested in the Liouville volume forms on $\mathcal{N}:=TN$ or $\mathcal{N}^*:=T^*N$, and their invariance under the spin-geodesic flow. The base point maps $\mathcal{N}\rightarrow N$ and $\mathcal{N}^*\rightarrow N$
will be denoted simply by $\pi$, or by $\pi^\mathcal{N}$ and $\pi^{\mathcal{N}^*}$,  if confusion with $N\rightarrow M$ may arise. 
The dimension of $N$ is
\[n:=d+\frac{d(d-1)}{2}=\frac{d(d+1)}{2},\]
where $d$ is the dimension of $M$.

The Kaluza-Klein  metric  on $N$ will be denoted simply as $\langle\cdot,\cdot\rangle$, or as $\langle\cdot,\cdot\rangle^N$ if it becomes necessary to be more explicit. This Riemannian metric induces the musical isomorphism $\flat: \mathcal{N}\rightarrow \mathcal{N}^*$.
The tautological $1$-form on the cotangent bundle $\mathcal{N}^*$, defined as \[\theta_\alpha(\zeta):=\alpha(d\pi_\alpha \zeta),\]  $\zeta\in T_\alpha\mathcal{N}^*$,  pulls back under $\flat$ to a $1$-form, still denoted $\theta$, on $\mathcal{N}$. 
Explicitly,
\[\theta_\xi(X):=\langle \xi, d\pi_\xi X\rangle,\]
for  $X\in T_\xi \mathcal{N}$. 
Due to $\flat$, we may restrict attention to $\mathcal{N}$.
The symplectic form on $\mathcal{N}$ is
\[\omega:=-d\theta\]
and the Liouville volume form is
\[\Omega:= \frac1{n!}\omega^n.\]

\begin{proposition}\label{prop:inner_prod_form_of_omega}
The exterior derivative $d\theta$ of the tautological one-form $\theta$ on $\mathcal{N}$ satisfies
\[d\theta_\xi(X_1, X_2)=\left\langle K_\xi X_1, d\pi_\xi X_2\right\rangle - \left\langle K_\xi X_2, d\pi_\xi X_1\right\rangle\]
for all $X_1, X_2\in T_\xi \mathcal{N}.$
\end{proposition}
\begin{proof}
Note that $\theta_\xi(X)=0$ whenever  $X$ is vertical. Thus, due to Proposition \ref{prop:bracket_relations},
\[d\theta\left(X_1^V, X_2^V\right)=X_1^V\theta\left( X_2^V\right)-X_2^V\theta\left( X_1^V\right) -\theta\left(\left[X_1^V,X_2^V\right]\right)=0. \]
If $\xi(t)$ is a parallel vector field along $\pi(\xi(t))\in N$ representing $X_1^H$ at $\xi=\xi(0)$,
\[X_1^H\theta\left(X_2^H\right)=\left.\frac{d}{dt}\right|_{t=0} \left\langle \xi(t), X_2\right\rangle = \langle \xi, \nabla_{X_1}X_2\rangle.\]
It follows that  
\begin{align*}
d\theta\left(X_1^H, X_2^H\right) &= X_1^H \theta\left(X_2^H\right)- X_2^H \theta\left(X_1^H\right) - \theta\left(\left[X_1^H, X_2^H\right]\right)\\
&=\langle \xi, \nabla_{X_1}X_2\rangle -\langle \xi, \nabla_{X_2}X_1\rangle -\langle \xi, [X_1, X_2]\rangle\\
&=0.
\end{align*}
Here we have used the fact that $\left[X_1^H, X_2^H\right]-[X_1, X_2]^H$ is a vertical vector, and that $\nabla$ is torsion-free. 
In similar fashion,
\[d\theta\left(X_1^V, X_2^H\right)=X_1^V\theta\left(X_2^H\right)-X_2^H\theta\left(X_1^V\right)-\theta\left(\left[X_1^V,X_2^H\right]\right)=X_1^V\theta\left(X_2^H\right).\]
Let $\xi(t)=\xi+tX_1$ represent  $X_1^V$, so that $d\pi_{\xi}\xi'(0)=0$ and $\frac{\nabla \xi}{dt}(0)=X_1$. Then
\[X_1^V\theta\left(X_2^H\right)=\left.\frac{d}{dt}\right|_{t=0}\langle \xi+t X_1, X_2\rangle =
\langle X_1, X_2\rangle=\left\langle K_\xi X_1^V, d\pi_{\xi}X_2^H\right\rangle.\]
The proposition  is a consequence of these observations and bilinearity of $d\theta$.
\end{proof}

  The geodesic spray $X^0$ is a Hamiltonian vector field on $\mathcal{N}$ associated to the Hamiltonian 
  \[H(S,\xi)=\frac12\|\xi\|^2_{S}.\] Therefore, 
  \[\mathcal{L}_{X^0}\omega = 0, \ \ \ \mathcal{L}_{X^0}\Omega=0.\]
  The $B$-perturbed vector field $X^B$ is not, in general, symplectic. To determine whether it preserves $\Omega$, it is sufficient to   investigate whether $\mathcal{L}_{Y^B}\Omega=0$. Note that
  \[\mathcal{L}_{Y^B}\omega = -d\left(i_{Y^B}d\theta\right)=-d\beta,\]
  where $\beta:=i_{Y^B}d\theta \in \Omega^1(\mathcal{N})$ satisfies (using Proposition \ref{prop:inner_prod_form_of_omega})
  \begin{equation}\label{eq:form_beta}\beta_{(S,\xi)}(Z)=\left\langle  K_\xi Y^B, d\pi_\xi Z\right\rangle_S=\left\langle  F_B(S,\xi), d\pi_\xi Z\right\rangle_S.\end{equation}
  As for the volume form, 
  \[\mathcal{L}_{X^B}\Omega= \frac1{(n-1)!}\left(\mathcal{L}_{X^B}\omega\right)\wedge \omega^{n-1}=-\frac1{(n-1)!}d\beta \wedge\omega^{n-1}.\]

  The calculations will be done locally. Let $E_1, \dots, E_d$ be a local orthonormal frame of vector fields with domain  $U\subseteq M$. From these we obtain on the set $\mathcal{U}\subseteq N$ of points projecting into $U$  a local orthonormal frame
  \[e_i(S) = E_i^H(S), \ \ e_{ij}(S)=S\circ (E_i(x)\wedge E_j(x)), \ \ i,j\in \{1,\dots, d\}, i<j.\]
And at  $\xi\in \pi^{-1}(\mathcal{U})\subseteq \mathcal{N}$ we define the horizontal (bars) and vertical (tildes) lifts
\begin{equation}\label{eq:bar_tilde_frame}\bar{e}_i(\xi),\ \  \tilde{e}_i(\xi),\ \  \bar{e}_{ij}(\xi), \ \ \tilde{e}_{ij}(\xi). \end{equation}
  These form a local orthonormal frame on $\pi^{-1}
(\mathcal{U})$ relative to the Sasaki metric on $\mathcal{N}$.  
We also introduce the dual frame
\begin{equation}\label{eq:bar_tilde_dual}\bar{\epsilon}_i(\xi),\ \  \tilde{\epsilon}_i(\xi),\ \  \bar{\epsilon}_{ij}(\xi), \ \ \tilde{\epsilon}_{ij}(\xi).\end{equation}
It is an immediate consequence of Proposition \ref{prop:inner_prod_form_of_omega} that
\[d\theta =\sum_{i} \tilde{\epsilon}_i\wedge \bar{\epsilon}_i +\sum_{i<j}\tilde{\epsilon}_{ij}\wedge \bar{\epsilon}_{ij}.\]
Here $\wedge$ is the usual wedge product of differential forms whereas $E_i\wedge E_j$ is the generalized cross-product.
The canonical volume form on $\mathcal{N}$ has, up to a multiplicative constant, the local form
\begin{equation}\label{eq:frame_expression_Omega}\Omega= C \tilde{\epsilon}_1 \wedge \bar{\epsilon}_1\wedge \cdots \wedge  \tilde{\epsilon}_d \wedge \bar{\epsilon}_d \bigwedge_{i<j} \tilde{\epsilon}_{ij}\wedge \bar{\epsilon}_{ij}.\end{equation}

We introduce the symmetric bilinear map $Q_S:T_SN\times  T_SN\rightarrow T_SN$ such that for $S\in N_x$ and  \[\xi_1 = (u_1, SW_1), \ \xi_2 = (u_2, SW_2)\in T_SN,\] 
we have
\begin{equation}\label{eq:Q}
\begin{aligned}Q_S(\xi_1,\xi_2)&:=\frac12 \left\{\left(-W_1B_xu_2-\mathcal{K}_S(u_2,W_1), (B_x u_1)\wedge u_2\right)\right. \\
&\qquad\qquad\qquad  \left.+\left(-W_2B_xu_1-\mathcal{K}_S(u_1,W_2), (B_x u_2)\wedge u_1\right)\right\}.\end{aligned}\end{equation}
Note that 
\[F_B(S,\xi)=Q_S(\xi, \xi). \]
We further introduce the functions
\begin{equation}\label{eq:QiQij}
Q_i(\xi) :=\left\langle Q_S(\xi,\xi), e_i(S)\right\rangle_S, \ \ Q_{ij}(\xi):=\left\langle Q_S(\xi,\xi), e_{ij}(S) \right\rangle_S.\end{equation}

\begin{lemma}\label{lemma:trace}
The following properties hold for $Q$:
\[\left\langle e_{ij}, Q(e_{ij},\xi)\right\rangle=0, \ \ \sum_i\left\langle e_i, Q(e_i,\xi)\right\rangle=0.\]
\end{lemma}
\begin{proof}
The first identity is an immediate consequence of the definition of $Q$. For the second,
\[\sum_i \left\langle e_i, Q(e_i,(u,SW))\right\rangle= 
-\frac12\text{Tr}\left(WB_x\right) -\frac12\sum_i\left\langle e_i, \mathcal{K}_S(e_i, W)\right\rangle. \]
The trace  of $WB_x$ is zero since $W^\intercal=-W$ and $B_x^\intercal=B_x$. 
Now recall that
\[\left\langle e_i, \mathcal{K}_S(e_i, W)\right\rangle =\left\langle W, \mathcal{C}_S(e_i, e_i)\right\rangle=0\]
since $\mathcal{C}$ is skew-symmetric.
\end{proof}

Note that \begin{equation}\label{eq:YB_expansion}
Y^B(\xi):=F_B(S,\xi)^V= \sum_{i} Q_i(\xi) \tilde{e}_i +\sum_{i<j} Q_{ij}(\xi)\tilde{e}_{ij}.
\end{equation}

\begin{lemma}\label{lemma:Lie_derivatives}
The following Lie derivative identities hold:
\begin{enumerate}
\item $\left(\mathcal{L}_{Y^B}\tilde{\epsilon}_i\right)\left(\tilde{e}_i\right)=2\left\langle e_i, Q(e_i,\xi)\right\rangle$;
\item $\left(\mathcal{L}_{Y^B}\bar{\epsilon}_i\right)\left(\bar{e}_i\right)=0$;
\item  $\left(\mathcal{L}_{Y^B}\tilde{\epsilon}_{ij}\right)\left(\tilde{e}_{ij}\right)=2\left\langle e_{ij}, Q(e_{ij},\xi)\right\rangle$;
\item $\left(\mathcal{L}_{Y^B}\bar{\epsilon}_{ij}\right)\left(\bar{e}_{ij}\right)=0$.
\end{enumerate}
\end{lemma}
\begin{proof}
First note that 
\[\left(\mathcal{L}_{Y^B}\tilde{\epsilon}_i\right)\left(\tilde{e}_i\right)=Y^B \tilde{\epsilon}\left(\tilde{e}_i\right)- \tilde{\epsilon}_i\left(\left[Y^B,\tilde{e}_i\right]\right)=-\left\langle \tilde{e}_i, \left[Y^B,\tilde{e}_i\right]\right\rangle.\]
Using the identity \eqref{eq:YB_expansion} and Proposition \ref{prop:bracket_relations}, we obtain
\[\left[Y^B, \tilde{e}_i\right]=-\sum_k\left(\tilde{e}_iQ_k\right)\tilde{e}_k - \sum_{k<\ell} \left(\tilde{e}_iQ_{k\ell}\right)\tilde{e}_{k\ell}.\]
Hence 
$\left(\mathcal{L}_{Y^B}\tilde{\epsilon}_i\right)\left(\tilde{e}_i\right)=\tilde{e}_iQ_i.$ For computing the derivative $\tilde{e}_iQ_i$,
let $\xi(t) =\xi +te_i(\pi(\xi))$ be a representative curve for $\tilde{e}_i(\xi)$. Clearly, $\pi(\xi(t))=\pi(\xi)$ is constant. Then, recalling the definition of $Q_i$ in \refeq{eq:QiQij}, we have
\[\left(\tilde{e}_i Q_i\right)(\xi)=\left.\frac{d}{dt}\right|_{t=0} \left\langle  e_i\left(\pi(\xi(t))\right), Q\left(\xi(t),\xi(t)\right)\right\rangle
= \left\langle  e_i\left(\pi(\xi) \right),  2Q_S\left(e_i(\pi(\xi)), \xi\right) \right\rangle.\]
This gives the first identity. The third identity is shown in a very similar fashion. For the second identity, one needs to evaluate the expression
\[\left(\mathcal{L}_{Y^B} \bar{\epsilon}\right)(\bar{e}_i)=-\left\langle \bar{e}_i, \left[Y^B, \bar{e}_i\right]\right\rangle,\]
which contains a Lie bracket of a vertical vector field and a horizontal vector field. This must be vertical by Proposition \ref{prop:bracket_relations}, hence the inner product is $0$. The last identity is zero for the same reason. 
\end{proof}

\begin{lemma}\label{lemma:Omega_YB_invariant}
The canonical volume form $\Omega$ on $\mathcal{N}$ satisfies $\mathcal{L}_{Y^B}\Omega=0$.
\end{lemma}
\begin{proof}
Using Equation (\ref{eq:frame_expression_Omega}) and the product rule for the Lie derivative with respect to the wedge product, we obtain
\[\mathcal{L}_{Y^B}\Omega = 2\left(\sum_i\left\langle e_i, Q(e_i,\xi) \right\rangle + \sum_{i<j}\left\langle e_{ij}, Q(e_{ij},\xi) \right\rangle \right)\Omega.\]
But the quantity between parentheses is zero by Lemma \ref{lemma:trace}. 
\end{proof}

\begin{theorem}\label{theor:main_vol_invariance}
The canonical volume form $\Omega$ on $\mathcal{N}$ is invariant under the spin-geodesic flow: $\mathcal{L}_{X^B}\Omega=0$.
\end{theorem}
\begin{proof}
We already noted above that $\mathcal{L}_{X^0}\Omega=0$ holds since $X^0$ (the geodesic spray) is a Hamiltonian vector field. 
Since $X^B=X^0+Y^B$, the claim is an immediate consequence of  Lemma \ref{lemma:Omega_YB_invariant}.
\end{proof}

\subsubsection{The reduced Liouville theorem}
We briefly describe volume preservation on the reduced phase space $\mathcal{A}$.
\begin{proposition}
The spin-geodesic vector field on the Atiyah algebroid \[\mathcal{A}\cong TM\oplus \mathfrak{so}(TM)\] preserves  the volume form
$\Omega_{\mathcal{A}}=\text{\em vol}_M\wedge \text{\em vol}_{\mathcal{A}/M}.$
\end{proposition}
\begin{proof}
Since divergence is a pointwise quantity, fix $x\in M$ and choose a local oriented orthonormal frame $E_1, \dots, E_d$ which is normal at $x$. 
Let $E_{ij}=E_i\wedge E_j$, $i<j$, be the associated orthonormal  frame of $\mathfrak{so}(TM)$, and write the fiber coordinates on $\mathcal{A}$ as
$(v_i, w_{ij})$. At $x$, the connection coefficients vanish and 
\[\Omega_{\mathcal{A}}=\text{vol}_M\, dv_1\cdots\,  dv_d \prod_{i<j}dw_{ij}.\]
The spin-geodesic equations are
\[\dot{x}=v, \ \ \ \frac{\nabla v}{dt} = -WBv, \ \ \ \frac{\nabla W}{dt}=(Bv)\wedge v.\]
Let $X^B_\mathcal{A}$ be the vector field on $\mathcal{A}$ defined by these equations.
Hence, at the chosen point, 
\[\text{div}_{\Omega_{\mathcal{A}}} X^B_{\mathcal{A}}= \text{Tr}_v\left(-WB\right) +\text{div}_W\left((Bv)\wedge v\right).\]
The second term vanishes because $(Bv)\wedge v$ is independent of $W$, while $\text{Tr}(WB)=0$ because $W$ is skew-symmetric and $B$ is symmetric. Therefore $\text{div}_{\Omega_{\mathcal{A}}} X^B_{\mathcal{A}}=0$ and we can conclude that the Lie derivative of
$\Omega_{\mathcal{A}}$ along $X^B_{\mathcal{A}}$ is zero.
\end{proof}

This is the reduced counterpart of Theorem \ref{theor:main_vol_invariance}. The proof on $N$ resolves the same cancellation relative to the orientation variable and the ordinary symplectic Liouville form. In particular, Lemma \ref{lemma:trace} is the corresponding trace calculation in the adapted orthonormal frame. 

\subsubsection{Invariant volume on level sets}
We show next that the restriction of the spin-geodesic flow on energy level manifolds in  $\mathcal{N}$ also leaves invariant volume forms.  
Let $n_H$ be the unit vector field on $\mathcal{N}$ defined by
\[d\pi_\xi n_H(\xi)=0, \ \ \ K_\xi n_H(\xi) =\frac{\xi}{\|\xi\|}.\]
\begin{lemma}\label{lemma:normal_field_nu}
The vector field $n_H$ on $\mathcal{N}$ is perpendicular to the level sets of the energy function $H(S,\xi)=\frac12\|\xi\|^2_{S}.$
\end{lemma}
\begin{proof}
Let $X$ be a vector field on $N$ and $\overline{X}, \widetilde{X}$ be its horizontal and vertical lifts to $\mathcal{N}$.
Let $\xi(t)$ be a curve in $\mathcal{N}$ representing $\overline{X}$ at $\xi$, hence $d\pi_\xi \xi'(0)=X$ and $\frac{\nabla^N\xi}{dt}(0)=0$.
Then
\[\left\langle  \text{grad}_\xi H, \overline{X}\right\rangle=\left.\frac{d}{dt}\right|_{t=0} H(\xi(t)) = \left.\frac{d}{dt}\right|_{t=0} \frac12 \|\xi(t)\|^2=\left\langle\xi,\frac{\nabla^N\xi}{dt}(0)\right\rangle=0.\]
Now let $\xi(t)$ represent $\widetilde{X}$ at $\xi$, hence $d\pi_\xi \xi'(0) =0$ and $K_\xi \xi'(0) = X(\pi(\xi))$. Then
\[\left\langle  \text{grad}_\xi H, \widetilde{X}\right\rangle= \left.\frac{d}{dt}\right|_{t=0} \frac12 \|\xi(t)\|^2=\left\langle\xi,\frac{\nabla^N\xi}{dt}(0)\right\rangle=\left\langle\xi,X\right\rangle=\|\xi\| \left\langle n_H, \widetilde{X} \right\rangle.\]
Therefore $ \text{grad}_\xi H - \|\xi\|  n_H(\xi)$ is perpendicular to all vectors in $T_\xi\mathcal{N}$, so it must be zero. This shows that
$n_H$ is collinear with the gradient of the kinetic energy function $H$, which is perpendicular to the energy level sets. 
\end{proof}
Given Lemma \ref{lemma:normal_field_nu}, the volume form induced on level sets of $H$ can be written as
\[\Omega^H=i_{n_H} \Omega.\]

\begin{lemma}\label{lemma:H_X0_theta}
The following properties hold.
\begin{enumerate}
\item $dH_\xi(X)=\left\langle \xi, K_\xi X\right\rangle$ for all $X\in T_\xi\mathcal{N}$;
\item $dH = -i_{X^0} d\theta$, thus $X^0$ is the symplectic gradient of $H$;
\item $\mathcal{L}_{X_0}\theta=dH$;
\item $\mathcal{L}_{X^0} d\theta = 0.$
\end{enumerate}
\end{lemma}
\begin{proof}
It follows immediately from Proposition \ref{prop:inner_prod_form_of_omega} that 
\[\left(i_{X^0}d\theta\right)_\xi(X)=d\theta_\xi(X^0, X)=-\left\langle \xi, K_\xi X\right\rangle.\] Using similar arguments as in that proposition (specifically, by choosing representative curves for $\xi(t)$ for vertical and horizontal lifts of vectors) we obtain
\[dH_\xi\left(\overline{X}\right) = \left.\frac{d}{dt}\right|_{t=0} \frac12 \left\langle \xi(t), \xi(t)\right\rangle = \left\langle \frac{\nabla^N\xi}{dt}(0),\xi(0) \right\rangle =0\]
and 
\[dH_\xi\left(\widetilde{X}\right)= \left\langle  \frac{\nabla^N\xi}{dt}(0), \xi\right\rangle =\left\langle \xi, K_\xi \widetilde{X}\right\rangle.\]
Therefore $dH = -i_{X^0} d\theta$ as claimed. 
Noting that $i_{X^0}\theta =\theta_\xi(X^0)=\|\xi\|^2=2H(\xi),$ we obtain
\[\mathcal{L}_{X^0}\theta = \left(di_{X^0}+i_{X^0}d\right)\theta = di_{X^0}\theta-dH =2dH -dH=dH,\]
which is the third assertion.
The last identity follows from the third since $\mathcal{L}_{X_0}$ commute with $d$. 
\end{proof}

We next show that $\mathcal{L}_{X^B}\Omega^H=0$. First observe that, by Lemma \ref{lemma:H_X0_theta} and Proposition \ref{prop:force_orthogonal_to_velocity},
\[dH(Y^B)=\left\langle \xi, K_\xi Y^B(\xi)\right\rangle=\left\langle \xi, K_\xi X^B(\xi)\right\rangle=\left\langle \xi, F_B(S,\xi)\right\rangle=0.\]
The same lemma also shows  that $dH(X^0)=0$, so $X^B$ is tangent to the level sets of $H$. Therefore,
the spin-geodesic  flow naturally restricts  to a flow on those level sets. 
\begin{theorem}\label{theor:volume_invariant_levels}   
The spin-geodesic flow restricts to a flow on positive  energy level sets and leaves invariant the volume form $\Omega^H$.
\end{theorem}
\begin{proof}
Let $Z(\xi):=\|\xi\| n_H(\xi)$. Note that $Z$ is a vertical vector field on $\mathcal{N}$ such that $K_\xi Z(\xi)=\xi$. Taking
 as in previous such arguments $\xi(t)$ a curve representing $Z(\xi)$, 
 \[dH_\xi(Z(\xi)) = \frac12 \left.\frac{d}{dt}\right|_{t=0} \left\langle \xi(t), \xi(t)\right\rangle= \left\langle K_\xi Z,\xi\right\rangle =2H(\xi).\]
 Thus $X^B dH(Z)=2 X^BH =0$ since $X^B$ is tangent to level sets of $H$. Therefore
 \[0=ddH(Z,X^B)=ZdH(X^B) -X^BdH(Z)-dH([Z,X^B]).\]
 This implies that $[Z,X^B]$ is also tangent to level sets of $H$. 
 In particular, the restriction of $i_{[X^B,Z]}\Omega$ to those level sets must vanish. Finally, note that
 \[\mathcal{L}_{X^B} \Omega^H = \mathcal{L}_{X^B}i_{n_H}\Omega = i_{n_H} \mathcal{L}_{X^B}\Omega + i_{[X^B, n_H]}\Omega = i_{[X^B,n_H]}\Omega.\]
 The restriction of the form $i_{[X^B,n_H]}\Omega$ to constant energy hypersurfaces also vanishes since $n_H$ differs from $Z$ multiplicatively by the function $\xi\mapsto \|\xi\|$, which is constant along those hypersurfaces.  
\end{proof}

\subsection{The form $\beta$ and non-symplecticity}\label{secion:non-symplectic}
We have seen that the condition for the spin-geodesic flow to preserve the symplectic form $\omega$ on $\mathcal{N}$ is that $d\beta=0$ where $\beta\in \Omega^1(\mathcal{N})$, defined in \eqref{eq:form_beta}, satisfies
\[\beta_{\xi}(Z)=\left\langle F_B(S,\xi), d\pi_\xi Z\right\rangle, \ \ \ \mathcal{L}_{Y^B}\omega = -d\beta.\]
Also recall the orthonormal frame \eqref{eq:bar_tilde_frame}  on $\mathcal{N}$ and its dual frame \eqref{eq:bar_tilde_dual} obtained from the horizontal and vertical lifts of a local orthonormal frame $e_i, e_{ij}$ on $N$. We use the index $a$ to designate either $i$ or the pair $(i,j)$. Now
write
\[F_a(\xi):=\left\langle F_B(S,\xi), e_a(S)\right\rangle.\]
 Note that these $F_a$ are the $Q_i, Q_{ij}$ previously introduced. Since $\beta$ annihilates vertical vectors of $TN\rightarrow N$, 
\[\beta= \sum_a F_a\bar{\epsilon}_a.\]
Associated to the local frame $e_a$ on $N$, let $\gamma_{ab}^{ c}$ be the locally defined functions on $N$ such that
\[[e_a, e_b]=\gamma_{ab}^{ c} e_c.\]
\begin{lemma}
With the notations just defined, we have
\begin{equation}\label{eq:dbeta_proposition}d\beta = \sum_{a,b} 2\left\langle Q(e_a,\xi), e_b\right\rangle \tilde{\epsilon}_a\wedge \bar{\epsilon}_b +\frac12\sum_{a,b}H_{ab}\bar{\epsilon}_a\wedge\bar{\epsilon}_b, \end{equation}
where 
\[H_{ab} =\bar{e}_a F_b -\bar{e}_b F_a  - \gamma_{ab}^c F_c.\]
In particular
\begin{equation}\label{eq:dbeta_tilde_bar}
d\beta\left(\tilde{e}_a,\bar{e}_b\right)=2\left\langle Q_S(e_a,\xi), e_b\right\rangle.
\end{equation}
\end{lemma}
\begin{proof}
There are three types of components of $d\beta$. 
The first   is
\[d\beta\left(\bar{e}_a, \bar{e}_b\right)=\bar{e}_a \beta(\bar{e}_b)-\bar{e}_b \beta(\bar{e}_a)-\beta([\bar{e}_a,\bar{e}_b])=\bar{e}_aF_b-\bar{e}_bF_a - \sum_{c}\gamma_{ab}^c F_c=H_{ab}.\]
Note that the curvature term  $[\bar{e}_a, \bar{e}_b]-[e_a,e_b]^H$ does not contribute because $\beta$ vanishes on vertical vectors.
The second type  is
\[d\beta\left(\tilde{e}_a,\bar{e}_b\right)=\tilde{e}_a\beta(\bar{e}_b)=\tilde{e}_aF_b =2\left\langle Q_S(e_a,\xi), e_b\right\rangle.\]
The third  is  
\[d\beta\left(\tilde{e}_a,\tilde{e}_b\right)=0,\]
due to $[\tilde{e}_a,\tilde{e}_b]=0$  (by Proposition \ref{prop:bracket_relations}).
Identity \eqref{eq:dbeta_proposition} is an immediate consequence of these remarks.
\end{proof}

Note that volume preservation only requires the trace of the block matrix with mixed coefficients $d\beta(\tilde{e}_a,\bar{e}_b)$ to vanish whereas
symplecticity  requires the entire two-form $d\beta$ to vanish. We call this matrix the {\em mixed block}.

Let us expand the  coefficients  in the mixed block. Let $E_1, \dots, E_d$ be a local  orthonormal frame in $M$. (Recall that $e_i$ is the horizontal lift of such $E_i$ to $N$) and set $E_{ij}=E_i\wedge E_j$. For $\xi=(v,W)\in T_SN$, the force is
\[F_B(S,\xi)=\left(-WBv-\mathcal{K}_S(v,W), (Bv)\wedge v\right).\]
From identity \eqref{eq:dbeta_tilde_bar} we obtain
\begin{align*}
d\beta(\tilde{e}_i, \bar{e}_k)&=-\left\langle WBE_i +\mathcal{K}_S(E_i,W), E_k\right\rangle, & d\beta(\tilde{e}_{i}, \bar{e}_{k\ell})&=\left\langle (BE_i)\wedge v + (Bv)\wedge E_i, E_{k\ell}\right\rangle\\
d\beta(\tilde{e}_{ij}, \bar{e}_{k})&=-\left\langle E_{ij}Bv+\mathcal{K}_S(v,E_{ij}), E_k \right\rangle, & d\beta(\tilde{e}_{ij}, \bar{e}_{k\ell})&=0.
\end{align*}
When $S=I_x$ (the identity isometry on $T_xM$) and $W=0$, these identities simplify (see Proposition \ref{prop:cal_C} for the vanishing of $\mathcal{C}_I$ and $\mathcal{K}_I$). In particular,
\[
d\beta(\tilde{e}_{ij}, \bar{e}_{k})=-\left\langle E_{ij}Bv, E_k \right\rangle=\left\langle Bv, E_{ij} E_k \right\rangle.\]
And if $i=k$, 
\[d\beta(\tilde{e}_{ij}, \bar{e}_{i})=\left\langle Bv, E_j \right\rangle\]
for all $v$ and $j$.
This shows that $d\beta$ is never identically zero so long as $B$ is not identically zero. 

Recall the tensor $C=\nabla^N-D$ introduced at the beginning of section \ref{sec:LC_connection_C}. 
It is identically zero if $M$ has dimension $2$ while,
for $d\geq 3$, $C\equiv 0$ is equivalent to $M$ being flat.
Clearly, if these tensors vanish,
$d\beta$ also vanishes. If $B$ vanishes but $C$ does not, 
\[d\beta(\tilde{e}_i, \bar{e}_k)=-\left\langle \mathcal{K}_S(E_i,W), E_k\right\rangle=-\left\langle W, C_S(E_i, E_k)\right\rangle. \]
If $C_S(E_i, E_k)\neq 0$, chose $W=C_S(E_i, E_k)$; then $d\beta(\tilde{e}_i, \bar{e}_k)\neq0$. So, symplecticity of  the spin-geodesic
flow requires vanishing of both $B$ and $C$.

We register these observations in the following theorem.
\begin{theorem}\label{theor:not_symplectic}
The spin-geodesic flow associated to a Riemannian manifold   of dimension $d\geq 2$ leaves invariant the Liouville volume for all spin-coupling fields $B$. On the other hand,  the flow preserves the symplectic form if and only if $B$ and $C$  are identically $0$. 
\end{theorem}
Thus nonzero spin coupling always destroys the canonical symplectic structure, while even at zero coupling, intrinsic curvature can obstruct symplecticity of the orientation-resolved flow in dimension $d\geq 3$. Nevertheless, the system is always conservative and preserves the canonical volume.

\end{document}